\documentclass[%
 preprint,
showpacs,
nofootinbib,
 amsmath,amssymb,
 prc,
 superscriptaddress
]{revtex4-1}

\usepackage{amsmath}
\usepackage{amssymb}
\usepackage{graphicx}% Include figure files
\usepackage{dcolumn}% Align table columns on decimal point
\usepackage{bm}% bold math
\usepackage{braket}
\usepackage{mathrsfs}
\usepackage{subcaption}

\begin{document}

%\preprint{APS/123-QED}

\title{
Resonant and Scattering States in the $\alpha+\alpha$ System from the Non-Localized Cluster Model
}% Force line breaks with \\


\author{Dong Bai}
\email{dbai@tongji.edu.cn}
%\email{dbai@alumni.itp.ac.cn}
\affiliation{School of Physics Science and Engineering, Tongji University, Shanghai 200092, China}%
\affiliation{School of Physics, Nanjing University, Nanjing 210093, China}%

\author{Zhongzhou Ren}
\email[Corresponding author: ]{zren@tongji.edu.cn}
\affiliation{School of Physics Science and Engineering, Tongji University, Shanghai 200092, China}%
\affiliation{Key Laboratory of Advanced Micro-Structure Materials, Ministry of Education, Shanghai 200092, China}



\begin{abstract}

The non-localized cluster model provides a new perspective on nuclear cluster effects and has been applied successfully to study cluster structures in various bound states and quasi-bound states (i.e., long-lived resonant states). In this work, we extend the application scope of the non-localized cluster model further to resonant and scattering states.
Following the $R$-matrix theory, the configuration space is divided into the interior and exterior regions by a large channel radius such that the nuclear forces and the antisymmetrization effects become negligible between clusters in the exterior region. In the interior region, the picture of non-localized clustering is realized mathematically by adopting the Brink-Tohsaki-Horiuchi-Schuck-R\"opke (Brink-THSR) wave functions as the bases to construct the interior wave functions. The Bloch-Schr\"odinger equation is used to match the interior wave functions continuously with the asymptotic boundary conditions of the resonant and scattering states at the channel radius, which leads eventually to solutions of the problem. As a first test of the formalism, the low-lying resonant states of ${}^{8}$Be and the phase shifts of the $\alpha+\alpha$ elastic scattering are studied. The numerical results agree well with the experimental data, which shows the validity of the theoretical framework.

\end{abstract}

%\pacs{21.10.Dr, 21.10.Gv, 03.75.Hh}% PACS, the Physics and Astronomy
\maketitle

%\tableofcontents

\section{Introduction}

Cluster structures are important for nuclear many-body problems and have been studied intensively by both experimentalists and theorists.
The non-localized cluster model is a new microscopic framework in nuclear cluster physics based on the picture of non-localized clustering \cite{Funaki:2015uya,Schuck:2017jtw,Tohsaki:2017hen,Freer:2017gip,Zhou:2019cjz}. It originates from the studies of $\alpha$-condensates by Tohsaki, Horiuchi, Schuck, and R\"opke (THSR) in 2001 \cite{Tohsaki:2001an}, and gets crystallized in the microscopic studies of ${}^{20}$Ne in 2012-2013 \cite{Zhou:2012zz,Zhou:2013ala,Zhou:2013eca}. In the traditional picture of localized clustering, the clusters are thought to be localized at fixed positions. Contrarily, in the picture of the non-localized clustering, the clusters could move freely in some nuclear containers. The non-localized cluster model has been applied to study nuclear structures of bound states and quasi-bound states (i.e., long-lived resonant states) in various light nuclei and hypernuclei, including ${}^{6}$He \cite{Lyu:2017ndp}, ${}^{8}$Be \cite{Funaki:2002fn,Funaki:2009fc}, ${}^{9}$Be \cite{Lyu:2014ewa}, ${}^{10}$Be \cite{Lyu:2015ika}, ${}^{11}$Be \cite{Lyu:2017zas}, ${}^{9}$B \cite{Zhao:2018roz}, ${}^{10}$B \cite{Zhao:2018nok}, ${}^{10}$C \cite{Zhao:2018nok}, ${}^{12}$C \cite{Funaki:2009fc,Suhara:2013csa,Tohsaki:2001an,Zhou:2014baa,Zhou:2016mhb,Itagaki:2018vnh,Zhou:2019hor}, ${}^{16}$O \cite{Funaki:2009fc,Suhara:2013csa,Tohsaki:2001an,Funaki:2017tia,Itagaki:2018vnh}, ${}^{20}$Ne \cite{Zhou:2012zz,Zhou:2013ala,Zhou:2017jhz}, ${}^{9}_\Lambda$Be \cite{Funaki:2014fba}, and ${}^{13}_{\,\,\Lambda}$C \cite{Funaki:2017asz}. The theoretical results agree well with the experimental data and the microscopic calculations based on the resonating group method (RGM) and the generator coordinate method (GCM),  revealing the robustness of the new picture.

In this work, we generalize the non-localized cluster model from bound and quasi-bound states to resonant and scattering states. Following the $R$-matrix theory \cite{Wigner:1946zz,Wigner:1946zz2,Wigner:1947zz,Lane:1948zh,Descouvemont:2010cx,Baye:1977vpg,Descouvemont:2012,Descouvemont:2015xoa}, the configuration space is divided into the interior and exterior regions. The channel radius separating these two regions has to be chosen properly such that in the exterior region the nuclear forces and the antisymmetrization effects become negligible  between different clusters and only the long-range Coulomb force survives. The Bloch-Schr\"odinger equation is adopted to match the interior wave functions continuously at the channel radius with the asymptotic boundary conditions of resonant and scattering states, which eventually leads to solutions of the problem.

In the interior region, the Brink-THSR wave functions \cite{Zhou:2013ala,Zhou:2013eca}, which combine features of the Brink wave functions \cite{Brink:1966} and the THSR wave functions \cite{Tohsaki:2001an}, are adopted as bases to construct the interior wave functions. The Brink wave functions are the canonical mathematical realizations of the localized clustering and assume the clusters to be localized at fixed generator coordinates. The THSR wave functions are, on the other hand, the canonical mathematical realizations of the non-localized clustering. For each THSR wave function, nuclear containers are introduced at the origin as extra ingredients to constrain the motion of clusters. Unlike the Brink wave functions, the clusters are assumed to be delocalized from any fixed positions and could move freely inside the nuclear containers. The Brink-THSR wave functions lie somewhere between the Brink and THSR wave functions. Compared with the THSR wave functions, the Brink-THSR wave functions have nuclear containers at different generator coordinates. The clusters then move non-locally inside these nuclear containers, which again contradicts the localized motion of the clusters in the Brink wave function. Therefore, the Brink-THSR wave functions could be regarded as another mathematical realizations of the non-localized clustering. Due to their rich hybrid structures, the Brink-THSR wave functions are shown previously to be crucial in describing the negative-parity states of ${}^{20}$Ne in the non-localized cluster model, which cannot be handled properly by starting from the THSR wave function directly \cite{Zhou:2013ala,Zhou:2013eca}. In other words, the Brink-THSR wave functions play the role of the ``midwife'' in establishing the new picture of non-localized clustering. Given these achievements, it is important to pursue further applications of the Brink-THSR wave functions. 

In the exterior region, the short-range nuclear forces between the clusters become negligible. So does the antisymmetrization effect between different clusters. These simplifications help determine the functional forms of the exterior wave functions. As to be shown later on, for the resonant states the relative components of the exterior wave functions are given by the outgoing Coulomb-Hankel functions, while for the scattering states the relative components of the exterior wave functions are given by combinations of the incoming and outgoing Coulomb-Hankel functions, with the relative coefficients given by the $S$-matrix elements. 

As a proof of concept, in this work we use the above theoretical formalism to study the resonant and scattering states in the $\alpha+\alpha$ system. The $\alpha+\alpha$ system has rich physical properties and is crucial for understanding many important nuclear reactions in astrophysics. 
%and has been studied by various experimental and theoretical groups. 
Both the low-lying resonances of ${}^{8}$Be and the phase shifts of the $\alpha+\alpha$ elastic scattering have been measured \cite{Abdullah:2006tww,Heydenburg:1956zza,Nilson:1958zz,Tombrello:1963,AFZAL:1969zz,Datar:2013pbd,Datar:2004sx,Tilley:2004zz}, making it an ideal playground to develop and validate our method. Various aspects of the $\alpha+\alpha$ system have been studied theoretically by many authors using the RGM \cite{Okai:1966zz,Thompson:1977zz}, the GCM \cite{Horiuchi:1970,Baye:1974dkx,Baye:1992zz,DohetEraly:2011zz}, the quantum Monte Carlo method \cite{Wiringa:2000gb,Wiringa:2013fia,Pastore:2014oda}, the THSR wave function \cite{Funaki:2002fn,Funaki:2009fc}, the cluster effective field theory \cite{Higa:2008dn,Andreatta:2020gmt}, the complex-scaled cluster model \cite{Kruppa:1988zz,Garrido:2012zx,Garrido:2013rta,Garrido:2013ewa}, the lattice effective field theory \cite{Elhatisari:2015iga}, the configuration interaction technique \cite{Kravvaris:2017nyj,Kravvaris:2019wva}, the $\delta$-shell potential method \cite{Luna:2019ufu}, etc. Also, the experience on studying the $\alpha+\alpha$ system would help extend our method further to the $\alpha+\alpha+\alpha$ system, which could contain more exotic structures such as gaslike $\alpha$-condensates \cite{Tohsaki:2001an}, linear-chain structures \cite{Suhara:2013csa}, etc.

The rest parts of this article are organized as follows: In Section \ref{TF}, we present the theoretical framework of our study, introducing briefly the non-localized cluster model in Subsection \ref{SubSectBTHSRWF} and the Bloch-Schr\"odinger equation in Subsection \ref{SubSectBSE}. The interaction model and the relevant matrix elements are given in Subsection \ref{ME}. In Section \ref{NR}, we present the numerical results on the low-lying resonances of ${}^{8}$Be and the phase shifts of the $\alpha+\alpha$ elastic scattering given by the non-localized cluster model and compare them with the experimental data. Section \ref{CC} ends this article with additional remarks and conclusions.

\section{Formalism}
\label{TF}

\subsection{Brink-THSR Wave Function}
\label{SubSectBTHSRWF}

%\subsection{Non-Localized Cluster Model and $R$-Matrix Theory}
%\label{THSR}

\begin{figure}
\centering

\begin{subfigure}[b]{\textwidth}
\centering
\minipage{0.5\textwidth}
  \includegraphics[width=0.85\linewidth]{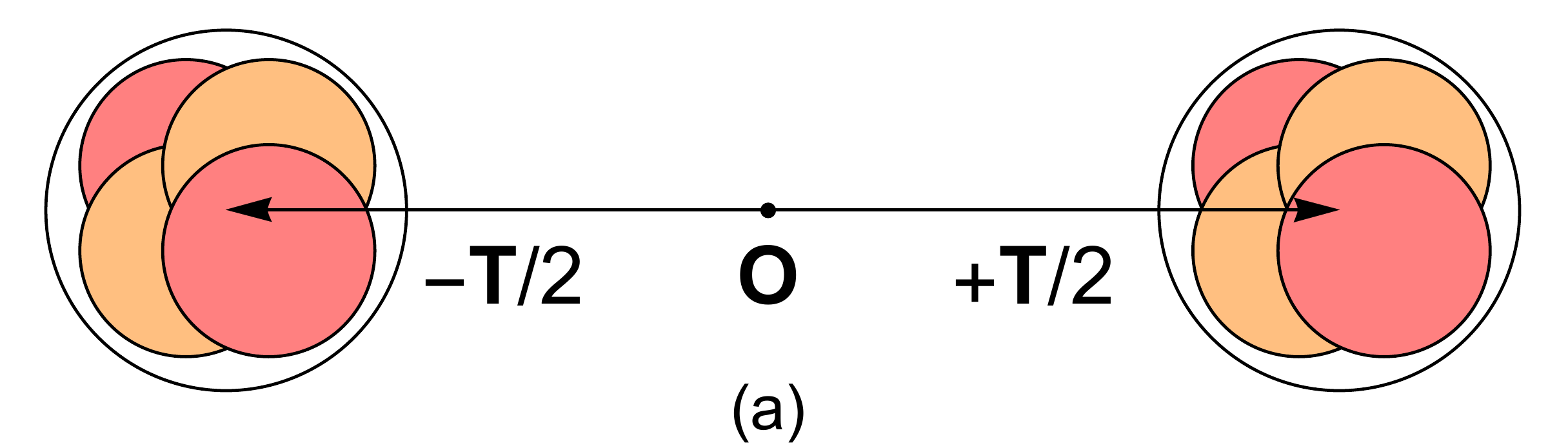}
%  \caption*{$\quad\ \ \ $(a)}
\endminipage\hfill
\minipage{0.5\textwidth}
  \includegraphics[width=\linewidth]{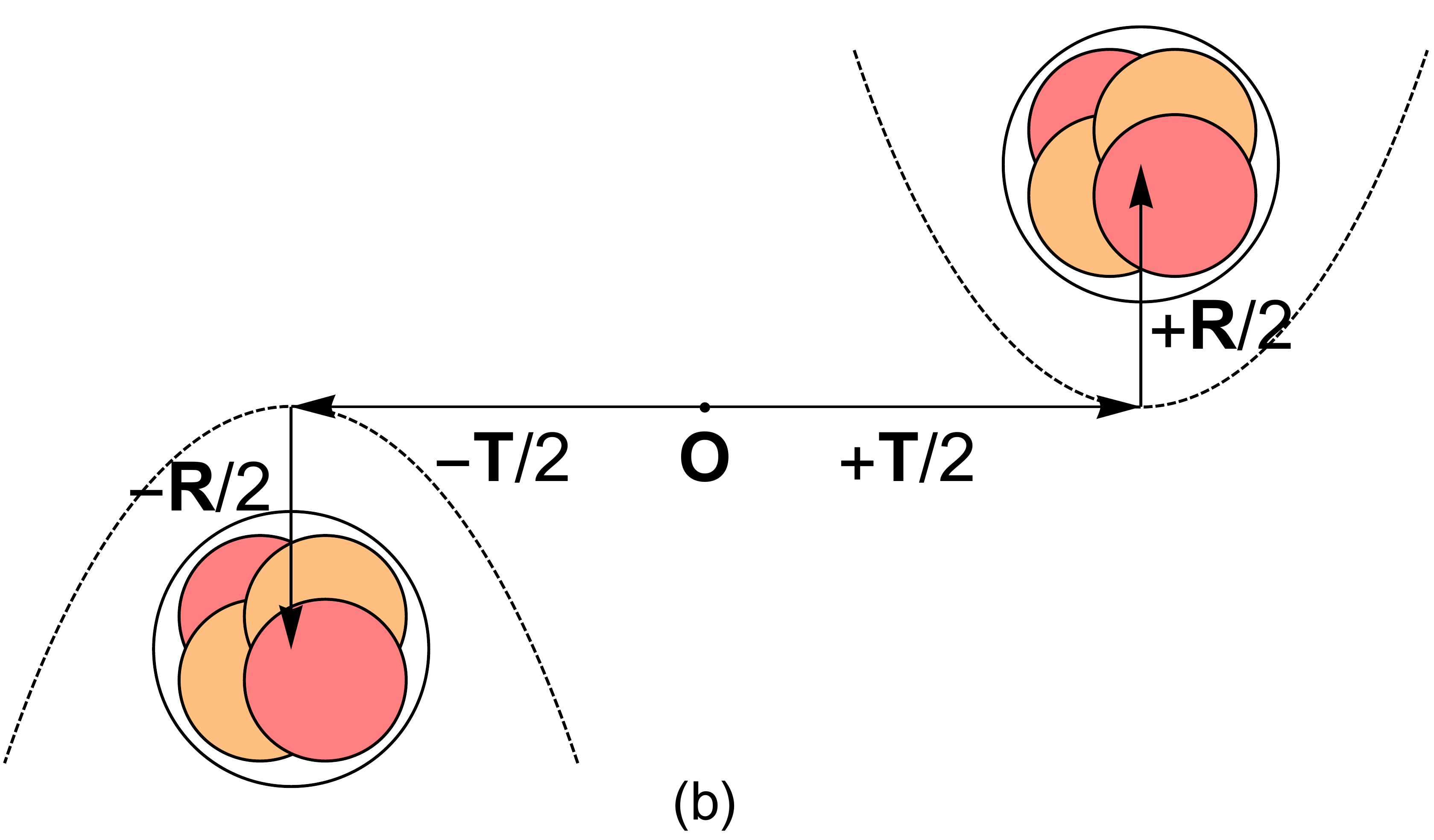}
%  \caption*{$\quad\ \ \ $(b)}
\endminipage
\end{subfigure}

\caption{(a) The $\alpha+\alpha$ system given by the Brink wave function, with two $\alpha$ clusters localized at the fixed positions $-\mathbf{T}/2$ and $+\mathbf{T}/2$. (b) The $\alpha+\alpha$ system given by the Brink-THSR wave function, with two $\alpha$ clusters moving non-locally inside the nuclear containers (dashed curves) fixed at $-\mathbf{T}/2$ and $+\mathbf{T}/2$.}
\label{Illus}
\end{figure}

We first present the theoretical formalism of the non-localized cluster model. It is adopted to describe the interior region of the $\alpha+\alpha$ system, where the antisymmetrization effect and nuclear interactions cannot be ignored safely and have to be handled exactly. The Brink-THSR wave functions are taken as the bases to construct the interior wave functions, with the expressions in the intrinsic frame given as follows:
%\begin{align}
%&\Phi_{2\alpha}(\beta)=\int\mathrm{d}^3R_1\mathrm{d}^3R_2\,\exp\!\left(-\frac{\mathbf{R}_1^2+\mathbf{R}_2^2}{\beta^2}\right)\Phi^\text{B}_{2\alpha}(\mathbf{R}_1,\mathbf{R}_2),\label{THSRWF}\\
%&\Phi^\text{B}_{2\alpha}(\mathbf{R}_1,\mathbf{R}_2)=\text{det}\{\varphi_{0s}(\mathbf{r}_1-\mathbf{R}_1)\chi_{\sigma_1\tau_1}\cdots\varphi_{0s}(\mathbf{r}_8-\mathbf{R}_2)\chi_{\sigma_8\tau_8}\},\\
%&\varphi_{0s}(\mathbf{r})=(\pi b^2)^{-3/4}\exp\left(-\frac{\mathbf{r}^2}{2b^2}\right),
%\end{align}
\begin{align}
&\Psi(\beta,\mathbf{T})=\mathscr{N}\int\mathrm{d}^3R\,\exp\!\left(-\frac{\mathbf{R}^2}{2\beta^2}\right)\Phi_\text{B}(\mathbf{R}+\mathbf{T}),\label{THSRWF}\\
&\Phi_\text{B}(\mathbf{R}+\mathbf{T})=\frac{1}{\sqrt{2}}\frac{1}{\sqrt{8!}}\text{det}\{\varphi_{0s}(\mathbf{r}_1-\mathbf{R}/2-\mathbf{T}/2)\chi_{\sigma_1\tau_1}\cdots\varphi_{0s}(\mathbf{r}_4-\mathbf{R}/2-\mathbf{T}/2)\chi_{\sigma_4\tau_4}\nonumber\\
&\qquad\qquad\qquad\qquad\qquad\ \times\varphi_{0s}(\mathbf{r}_5+\mathbf{R}/2+\mathbf{T}/2)\chi_{\sigma_5\tau_5}\cdots\varphi_{0s}(\mathbf{r}_8+\mathbf{R}/2+\mathbf{T}/2)\chi_{\sigma_8\tau_8}\},\label{BWF}\\
&\varphi_{0s}(\mathbf{r}\pm\mathbf{R}/2\pm\mathbf{T}/2)=(\pi b^2)^{-3/4}\exp\left[-\frac{(\mathbf{r}\pm\mathbf{R}/2\pm\mathbf{T}/2)^2}{2b^2}\right].
\end{align}
Here, $\varphi_{0s}(\mathbf{r})$ and $\chi_{\sigma\tau}$ are the spatial and spin-isospin wave functions of a single nucleon. $\Phi_\text{B}(\mathbf{R}+\mathbf{T})$ is the Brink wave function with $\mathbf{R}+\mathbf{T}$ being the generator coordinate and can be interpreted intuitively as two $\alpha$ clusters at the fixed positions $-(\mathbf{R}+\mathbf{T})/2$ and $+(\mathbf{R}+\mathbf{T})/2$. The factor $1/\sqrt{2}$ in Eq.~\eqref{BWF} accounts for the indistinguishability of the two $\alpha$ clusters. In the Brink-THSR wave function, the weight function for the generator coordinate $\mathbf{R}$ is taken to be the Gaussian function with the width parameter given by $\beta$. For the real $\beta$ , the Brink-THSR wave function could be interpreted intuitively as two $\alpha$ clusters moving non-locally inside two nuclear containers located at the fixed positions $-\mathbf{T}/2$ and $+\mathbf{T}/2$, with the container sizes determined by $\beta$. A pictorial illustration of both the Brink wave function $\Phi_\text{B}(\mathbf{T})$ and the Brink-THSR wave function $\Psi(\beta,\mathbf{T})$ could be found in Fig.~\ref{Illus}. The overall normalization constant in Eq.~\eqref{THSRWF} is chosen to be $\mathscr{N}=1/(2\pi\beta^2)^{3/2}$.
%, which allows the Brink-THSR wave function $\Psi(\beta,\mathbf{T})$ to be reduced exactly to the Brink wave function $\Phi_\text{B}(\mathbf{T})$ in the limit of $\beta\to0$. 
%It is helpful to notice that, besides real numbers, the parameter $\beta$ could also take pure imaginary values such that $b^2+2\beta^2>0$. 
Thanks to the analytic solvability of the Gaussian integration, Eq.~\eqref{THSRWF} could be further simplified:
\begin{align}
&\Psi(\beta,\mathbf{T})=%\left(\frac{8}{\pi b^2}\right)^{3/4}
\Psi_\text{CM}(\mathbf{X}_\text{CM})\times\widehat{\Psi}(\beta,\mathbf{T}),\label{COMSep}\\
&\Psi_\text{CM}(\mathbf{X}_\text{CM})=\left(\frac{8}{\pi b^2}\right)^{3/4}\!\!\!\!\exp\!\left({-\frac{4\mathbf{X}^2_\text{CM}}{b^2}}\right),\label{COMSep2}\\
&\widehat{\Psi}(\beta,\mathbf{T})=\frac{1}{\sqrt{140}}\,\mathscr{A}_{12}\left[\Gamma(\bm{\rho},\beta,\mathbf{T})\widehat{\phi}(\alpha_1)\widehat{\phi}(\alpha_2)\right]\label{ITHSR},\\
&\Gamma(\bm{\rho},\beta,\mathbf{T})=\left(\frac{2}{\pi}\right)^{3/4}\!\!\!\!\frac{b^{3/2}}{(b^2+2\beta^2)^{3/2}}\exp\!\left[-\frac{(\bm{\rho}-\mathbf{T})^2}{b^2+2\beta^2}\right],\label{GTHSR}
%&\bm{\rho}=\mathbf{X}_1-\mathbf{X}_2,
\end{align}
where $\bm{\rho}=\mathbf{X}_1-\mathbf{X}_2$ is the relative coordinate and $\mathbf{X}_\text{CM}=\frac{1}{2}(\mathbf{X}_1+\mathbf{X}_2)$ is the center-of-mass (CM) coordinate of the $\alpha+\alpha$ system, with $\mathbf{X}_i=\frac{1}{4}\sum_{j=4i-3}^{4i}\mathbf{r}_{j}\ (i=1,2)$ being the CM coordinate of the $i$th $\alpha$ cluster, and $\widehat{\phi}(\alpha_i)$ is the antisymmetrized and normalized internal wave function of the $i$th $\alpha$ cluster and is connected to the Brink wave function by
\begin{align}
&\frac{1}{\sqrt{4!}}\text{det}\left\{\varphi_{0s}(\mathbf{r}_{4i-3}+(-1)^{i}(\mathbf{R}/2+\mathbf{T}/2))\chi_{\sigma_{4i-3}\tau_{4i-3}}\cdots\varphi_{0s}(\mathbf{r}_{4i}+(-1)^{i}(\mathbf{R}/2+\mathbf{T}/2))\chi_{\sigma_{4i}\tau_{4i}}\right\}\nonumber\\
=&\left(\frac{4}{\pi b^2}\right)^{3/4}\exp\left\{-\frac{2}{b^2}\left[\mathbf{X}_i+(-1)^{i}(\mathbf{R}/2+\mathbf{T}/2)\right]^2\right\}\widehat{\phi}(\alpha_i).
\end{align}
The intercluster antisymmetrization operator $\mathscr{A}_{12}$ in Eq.~\eqref{ITHSR} is defined as
\begin{align}
\mathscr{A}_{12}=1-\sum_{\substack{i\in\alpha_1\\j\in\alpha_2}}\mathscr{P}_{ij}+\cdots,
\end{align}
where $\mathscr{P}_{ij}$ exchanges the $i$th nucleon in $\alpha_1$ with the $j$th nucleon in $\alpha_2$, etc.
% and the overall factor $\frac{1}{{2}}$ takes care of the indistinguishability of the two $\alpha$ clusters.
The Brink-THSR wave function has the merit to have the CM motion be easily separated out and captured by the normalized wave function $\Psi_\text{CM}(\mathbf{X}_\text{CM})$ in Eqs.~\eqref{COMSep} and \eqref{COMSep2}. 
%In the center-of-mass frame with $\mathbf{X}_\text{CM}=0$, we have $\Psi_{2\alpha}(\beta)=\widehat{\Phi}_{2\alpha}(\beta)$. 
%The parameter $B=\sqrt{b^2+2\beta^2}$ in Eq.~\eqref{GTHSR} describes the size of the nuclear container within which the two $\alpha$ clusters move. 

To describe physical states with the definite angular momentum and parity, we consider further the partial-wave expansion of the Brink-THSR wave function
\begin{align}
&\Psi(\beta,\mathbf{T})=\Psi_\text{CM}(\mathbf{X}_\text{CM})\times4\pi\sum_{LM}\widehat{\Psi}_{L}(\beta,T)Y_{LM}(\Omega_\rho)Y^*_{LM}(\Omega_T),\\
&\widehat{\Psi}_{L}(\beta,T)=\frac{1}{\sqrt{140}}\mathscr{A}_{12}\Gamma_L(\rho,\beta,T)\widehat{\phi}(\alpha_1)\widehat{\phi}(\alpha_2),\\
&\Gamma_L(\rho,\beta,T)=\left(\frac{2}{\pi}\right)^{3/4}\!\!\!\!\frac{b^{3/2}}{(b^2+2\beta^2)^{3/2}}\exp\!\left(-\frac{\rho^2+T^2}{b^2+2\beta^2}\right)\text{i}_L\left(\frac{2\rho T}{b^2+2\beta^2}\right).
\end{align}
Here, $\text{i}_L(x)=\sqrt{\frac{\pi}{2x}}I_{L+1/2}(x)$, with $I_{L+1/2}(x)$ being the modified Bessel function of the first kind.
Then, the radial component of the interior wave function $\widehat{\Psi}^\text{int}_{L}(E)$ at the reaction energy $E$ (in the CM frame) could be given by
\begin{align}
\widehat{\Psi}^\text{int}_{L}(E)&=\int\mathrm{d}T f_L(T,E) \widehat{\Psi}_L(\beta,T)=\sum_n \widetilde{f}_L(T_n,E) \widehat{\Psi}_L(\beta,T_n),\label{IWF}
%&\equiv\mathscr{A}\phi(\alpha_1)\phi(\alpha_2)g_\text{int}(\rho)\\
%&\to \phi(\alpha_1)\phi(\alpha_2)g_\text{int}(\rho), \qquad \rho\to\infty
\end{align}  
with $f_L(T,E)$ being the weight function and $\left\{\widetilde{f}_L(T_n,E)\right\}$ being the corresponding discretized representation.
%where $E$ is the reaction energy in the center-of-mass frame, and $g_\text{int}(\rho)$ is the relative wave function in the internal region.

\subsection{Bloch-Schr\"odinger Equation}
\label{SubSectBSE}

Following the $R$-matrix theory, the channel radius $a$ separates the interior and exterior regions and is chosen to be so large that the short-range nuclear interaction and the antisymmetrization could be safely neglected between the two $\alpha$ clusters in the exterior region. Therefore, in the exterior region the Hamiltonian becomes 
\begin{align}
&H_L\rightarrow H_L^\text{ext}\equiv H_{\alpha_1}+H_{\alpha_2}+T_{{\rho}}+\frac{Z_\alpha^2e^2}{\rho},\\
&T_{{\rho}}=\frac{\hbar^2}{2\mu}\left[-\frac{1}{\rho^{2}}\frac{\partial}{\partial\rho}\left(\rho^{2}\frac{\partial}{\partial\rho}\right)+\frac{L(L+1)}{\rho^{2}}\right].
\end{align}
with $H_{\alpha_1}$ and $H_{\alpha_2}$ being the intrinsic Hamiltonian of the two $\alpha$ clusters.
The radial component of the exterior wave function takes the following form for the resonant and scattering states, respectively
\begin{align}
&\widehat{\Psi}^\text{ext}_L(E)=\frac{1}{\sqrt{35}}g^\text{ext}_L(\rho)\widehat{\phi}(\alpha_1)\widehat{\phi}(\alpha_2),\label{EPsi}\\
&g^\text{ext}_L(\rho)=
\begin{cases}
\\[-6ex]
&\!\!\!\!\mathcal{H}^{(+)}_L(\eta,k\rho)/{\rho}, \quad \text{for resonant states}\\
&\!\!\!\!\left[\mathcal{H}^{(-)}_L(\eta,k\rho)-\mathcal{S}_L(E) \mathcal{H}^{(+)}_L(\eta,k\rho)\right]/{\rho}, \quad \text{for scattering states}\label{EWF}
\end{cases}
\end{align}
where $\mathcal{H}^{(\mp)}_L(\eta,k\rho)$ are the incoming/outgoing Coulomb-Hankel functions, with $k=\frac{\sqrt{2\mu E}}{\hbar}$ being the wave number, $\eta=\frac{Z_\alpha^2e^2}{\hbar}\sqrt{\frac{\mu}{2E}}$ being the Coulomb-Sommerfeld parameter, and $\mu$ being the two-body reduced mass. $\mathcal{S}_L(E)$ is the so-called $S$-matrix element and is related to the phase shift $\delta_L(E)$ by
%\begin{align}
$\mathcal{S}_L(E)=\exp(2i \delta_L(E))$. 
%\end{align}
The exterior wave function for the resonant state in Eq.~\eqref{EWF} needs some remarks. In this work, we follow Siegert and define the resonant states in the framework of non-Hermitian quantum mechanics as eigenstates with purely outgoing asymptotes \cite{Moiseyev:2011}. The eigenvalues of the resonant states are given by complex numbers ${E}=\mathcal{E}-i\Gamma/2$, with $\mathcal{E}$ being the energy and $\Gamma$ being the decay width. 
%Compared with the usual Hermitian quantum mechanics, it is important to note that, the scalar product in the non-Hermitian quantum mechanics is defined by
%\begin{align}
%&\braket{g|f}=\int g(\mathbf{r})f(\mathbf{r})\mathrm{d}^3r,
%\end{align}
%for the non-Hermitian quantum mechanics, instead of $\braket{g|f}=\int g(\mathbf{r})^*f(\mathbf{r})\mathrm{d}^3r$ for the Hermitian quantum mechanics. When $f(\mathbf{r})$ and $g(\mathbf{r})$ are real functions, these two definitions of scalar products coincide with each other. For comprehensive discussions on the non-Hermitian quantum mechanics, we recommend  the textbook by N.~Moiseyev \cite{Moiseyev:2011}.

Given the interior and exterior wave functions in Eqs.~\eqref{IWF} and \eqref{EPsi}, the coefficients $\left\{\widetilde{f}_L(T_n,E)\right\}$
%and the $S$ matrix $\mathcal{S}_l(E)$ %The continuity condition of the wave function requires that 
%\begin{align}
%$g_\text{int}(a)=g_\text{out}(a)$
%\end{align}
%at the channel radius $a$. Moreover, 
can be determined by solving the Bloch-Schr\"odinger equation \cite{Bloch:1957}
\begin{align}
(H_L+\mathcal{L}(B)-E)\Psi^\text{int}_L=\mathcal{L}(B)\Psi^\text{ext}_L.\label{BSE}
\end{align}
The Bloch operator $\mathcal{L}(B)$ gives an elegant implementation of the continuity condition at the channel radius and is given explicitly by
\begin{align}
\mathcal{L}(B)=35\frac{\hbar^2}{2\mu a}\delta(\rho-a)\left(\frac{\mathrm{d}}{\mathrm{d}\rho}\rho-B\right).\label{DefL}
\end{align}
Here, the parameter $B$ could take arbitrary values. The prefactor $35=\frac{8!}{2\times4!4!}$ is the number of equivalent definitions of the relative coordinate $\bm{\rho}$.
%for the $S$-wave scattering. Here, $l$ counts all the possible linearly independent relative coordinates, and it is this summation over $l$ in Eq.~\eqref{DefL} that helps preserve the permutation symmetry of the modified Hamiltonian. 
Substituting Eq.~\eqref{IWF} into Eq.~\eqref{BSE}, we have
\begin{align}
&\sum_{n'} [C(B,E)]_{nn'} \widetilde{f}_L(T_{n'},E) = \braket{\widehat{\Psi}_L(\beta,T_n)|\mathcal{L}(B)|\widehat{\Psi}_L^\text{ext}(E)},\label{LBSE}\\
&[C(B,E)]_{nn'}=\left(\widehat{\Psi}_L(\beta,T_n)|H_L+\mathcal{L}(B)-E|\widehat{\Psi}_L(\beta,T_{n'})\right).\label{CMatrix}
\end{align}
The round brackets ``(\ )'' in Eq.~\eqref{CMatrix} refer to the interior matrix element, which is evaluated within the interior region only. For the resonant states, we take
\begin{align}
B=B_{*}\equiv ka\frac{\mathcal{H}^{(+)'}_L(\eta,ka)}{\mathcal{H}^{(+)}_L(\eta,ka)},
\end{align}
such that the right-hand side of Eq.~\eqref{LBSE} vanishes. Here, $\mathcal{H}^{(\mp)'}_L(\eta,ka)$ is the derivative of $\mathcal{H}^{(\mp)}_L(\eta,ka)$ with respect to $ka$. The energy spectrum of the resonant states could then be obtained by solving the following generalized eigenvalue problem
\begin{align}
\sum_{n'}\left(\widehat{\Psi}_L(\beta,T_n)|H_L+\mathcal{L}(B_*)|\widehat{\Psi}_L(\beta,T_{n'})\right)\widetilde{f}_L(T_{n'},E)=E\sum_{n'}\left(\widehat{\Psi}_L(\beta,T_n)|\widehat{\Psi}_L(\beta,T_{n'})\right)\widetilde{f}_L(T_{n'},E).\label{GEP4RS}
\end{align}
Noticeably, the parameter $B_*$ depends implicitly on the energy $E$ through the definition of the wave number $k$. Therefore, Eq.~\eqref{GEP4RS} has to be solved in a self-consistent manner, i.e., one starts with some well-guessed values of $E$ and iterates until the numerical results converge.
For the scattering states, we take $B=0$ for simplicity. With the matrix elements $\{[C(0,E)]_{nn'}\}$, the $R$- and $S$-matrix elements are given by
\begin{align}
&\mathcal{R}_L=\frac{\hbar^2a}{2\mu}\sum_{nn'}\Gamma_L(a,\beta,T_n)[C(0,E)]^{-1}_{nn'}\Gamma_{L}(a,\beta,T_{n'}),\\
&\mathcal{S}_L=\frac{\mathcal{H}^{(-)}_L(\eta,ka)-ka\,{\mathcal{H}^{(-)'}_L}(\eta,ka)\mathcal{R}_L}{\mathcal{H}^{(+)}_L(\eta,ka)-ka\,\mathcal{H}^{(+)'}_L(\eta,ka)\mathcal{R}_L}.\label{SMat}
%\\
%&\Gamma_{2\alpha}(\rho,\beta)\equiv\left(\frac{\pi B^2}{2}\right)^{-3/4}\exp\left(-\frac{\bm{\rho}^2}{B^2}\right), \quad\text{with } B=\sqrt{b^2+2\beta},
\end{align}
The phase shifts could be obtained from its definition. With the $S$-matrix element given in Eq.~\eqref{SMat}, the interior wave function $\Psi^\text{int}_L$ could be obtained by solving the linear equations given by Eq.~\eqref{LBSE}. 

%Following Ref.~\cite{}, we study further the container evolution in the $\alpha$-$\alpha$ elastic scattering by fitting the internal wave function $\Psi_\text{int}$ with a single THSR wave function $\widehat{\Phi}_{2\alpha}(\beta)$, which could be done by calculating the squared overlap
%\begin{align}
%\mathcal{O}(\beta,E)=\frac{\left|\left(\widehat{\Phi}_{2\alpha}(\beta)|\Psi_\text{int}\right)\right|^2}{\left(\widehat{\Phi}_{2\alpha}(\beta)|\widehat{\Phi}_{2\alpha}(\beta)\right)\times\left(\Psi_\text{int}|\Psi_\text{int}\right)},
%\end{align} 
%and studying its landscape.

\subsection{Interaction Model and Interior Matrix Elements}
\label{ME}

The microscopic Hamiltonian for the $\alpha+\alpha$ system is given by
\begin{align}
&H=T-T_\text{CM}+V_N+V_C,\\
&T-T_\text{CM}=-\sum_{i=1}^8\frac{\hbar^2}{2m}\left(\frac{\partial}{\partial\mathbf{r}_i}\right)^2+\frac{\hbar^2}{16m}\left(\frac{\partial}{\partial\mathbf{X}_\text{CM}}\right)^2,\\
&V_N=\frac{1}{2}\sum_{i\neq j}^8\sum_{k=1}^{N_g}V_{k}\exp(-(r_{ij}/a_k)^2)(w_k-m_kP^\sigma_{ij}P^\tau_{ij}+b_kP^\sigma_{ij}-h_kP^\tau_{ij}),\\
&V_C=\frac{1}{2}\sum_{i\neq j}^8\frac{e^2}{r_{ij}}\left(\frac{1}{2}+t_{iz}\right)\left(\frac{1}{2}+t_{jz}\right),
\end{align}
where $N_g$ is the number of the Gaussian form factors used in the effective nucleon-nucleon central interaction, $P^{\sigma}_{ij}$ and $P^\tau_{ij}$ are the spin and isospin exchange operators, and the isospin $z$-component equals $t_{z}=+1/2$ for the proton and $t_{z}=-1/2$ for the neutron.

The interior matrix elements could be calculated by subtracting the exterior contributions from the whole-space matrix elements. Explicitly, we have
\begin{align}
&\left(\widehat{\Psi}_{L}(\beta,T_n)|\widehat{\Psi}_{L}(\beta,T_{n'})\right)=\braket{\widehat{\Psi}_L(\beta,T_n)|\widehat{\Psi}_L(\beta,T_{n'})}-\int_{a}^\infty\!\!\!\!\mathrm{d}\rho\,\rho^2\, \Gamma_{L}({\rho},\beta,T_n)\Gamma_{L}({\rho},\beta,T_{n'}),\label{IntOver}\\%\nonumber\\
%&=\braket{\Psi_{2\alpha}(\beta_n)|H|\Psi_{2\alpha}(\beta_{n'})}-4\pi^2c\int_{a}^\infty\!\mathrm{d}\rho\,\rho^2\,\Gamma_{2\alpha}(\bm{\rho},\beta_n)\left(2E_\alpha+T_\rho+\frac{Z_\alpha^2e^2}{\rho}\right)\Gamma_{2\alpha}(\bm{\rho},\beta_{n'}),%\nonumber\\
%\end{align}
%for the overlap of THSR wave functions, and
%\begin{align}
&\left(\widehat{\Psi}_{L}(\beta,T_n)|H_L|\widehat{\Psi}_{L}(\beta,T_{n'})\right)=\braket{\widehat{\Psi}_{L}(\beta,T_n)|H_L|\widehat{\Psi}_{L}(\beta,T_{n'})}\nonumber\\
&\qquad\qquad\qquad\qquad\qquad\quad\, -\int_{a}^\infty\!\!\!\!\mathrm{d}\rho\,\rho^2\, \Gamma_{L}({\rho},\beta,T_n)H_L^\text{ext}\Gamma_{L}({\rho},\beta,T_{n'}),\label{IntHami}\\
%&=\braket{\Psi_{2\alpha}(\beta_n)|H|\Psi_{2\alpha}(\beta_{n'})}-4\pi c\int_{a}^\infty\!\mathrm{d}\rho\,\rho^2\,\Gamma_{2\alpha}(\bm{\rho},\beta_n)\left(2E_\alpha+T_\rho+\frac{Z_\alpha^2e^2}{\rho}\right)\Gamma_{2\alpha}(\bm{\rho},\beta_{n'}),%\nonumber\\
%\end{align}
%where $E_\alpha$ is the binding energy of the $\alpha$ particle. The matrix element of the Bloch operator is given by
%\begin{align}
&\left(\widehat{\Psi}_{L}(\beta,T_n)|\mathcal{L}(B)|\widehat{\Psi}_{L}(\beta,T_{n'})\right)%\nonumber\\
%&=4\pi c\int_0^\infty\mathrm{d}\rho\,\rho^2\,\Gamma_{2\alpha}(\bm{\rho},\beta_n)\frac{\hbar^2}{2\mu a}\delta(\rho-a)\frac{\mathrm{d}}{\mathrm{d}\rho}(\rho\Gamma_{2\alpha}(\bm{\rho},\beta_{n'}))\nonumber\\
%&=4\pi c \frac{\hbar^2}{2\mu a} \int_0^\infty\mathrm{d}\rho\,\rho^2\,\Gamma_{2\alpha}(\bm{\rho},\beta_n)\delta(\rho-a)(\Gamma_{2\alpha}(\bm{\rho},\beta_{n'})+\rho\frac{\mathrm{d}}{\mathrm{d}\rho}\Gamma_{2\alpha}(\bm{\rho},\beta_{n'}))\nonumber\\
= \frac{\hbar^2a}{2\mu}\,\Gamma_{L}(a,\beta,T_n)\left[\Gamma_{L}(a,\beta,T_{n'})+a\frac{\mathrm{d}}{\mathrm{d}a}\Gamma_{L}(a,\beta,T_{n'})\right]\nonumber\\
&\qquad\qquad\qquad\qquad\qquad\quad\ \ \ \ -\frac{\hbar^2a}{2\mu}B\,\Gamma_L(a,\beta,T_n)\Gamma_L(a,\beta,T_{n'}),\\
&\left(\widehat{\Psi}_{L}(\beta,T_n)|\mathcal{L}(B)|\widehat{\Psi}^\text{ext}_{L}(E)\right)
= \frac{\hbar^2a}{2\mu}\,\Gamma_{L}(a,\beta,T_n)\left[g^\text{ext}_{L}(a)+a\frac{\mathrm{d}}{\mathrm{d}a}g^\text{ext}_{L}(a)\right]\nonumber\\
&\qquad\qquad\qquad\qquad\qquad\ \ \ \ \ -\frac{\hbar^2a}{2\mu}B\,\Gamma_L(a,\beta,T_n)g^\text{ext}_L(a).
\end{align}
The whole-space matrix elements in Eqs.~\eqref{IntOver} and \eqref{IntHami} could be evaluated by using
\begin{align}
%&\quad\braket{\Phi(\mathbf{R})|\Phi(\mathbf{R}')}\nonumber\\
%&=4\pi\sum_{l}(2l+1)\braket{\widehat{\Phi}_l(R)|\widehat{\Phi}_l(R')}P_l(\cos\theta).
&\braket{\widehat{\Psi}_L(\beta,T)|\widehat{\Psi}_L(\beta,T')}=\frac{1}{8\pi}\int_0^\pi\braket{\Psi(\beta,\mathbf{T})|\Psi(\beta,\mathbf{T}')}P_L(\cos\theta)\,\sin\theta\,\mathrm{d}\theta,\\
&\braket{\widehat{\Psi}_L(\beta,T)|H_L|\widehat{\Psi}_L(\beta,T')}=\frac{1}{8\pi}\int_0^\pi\braket{\Psi(\beta,\mathbf{T})|H|\Psi(\beta,\mathbf{T}')}P_L(\cos\theta)\,\sin\theta\,\mathrm{d}\theta.%,\\
%&\braket{\Phi(\mathbf{R})|\Phi(\mathbf{R}')}=\\
%&\braket{\Phi(\mathbf{R})|H|\Phi(\mathbf{R}')}=.
\end{align}
Here, we take the generator coordinate $\mathbf{T}$ to be along the $z$ axis and $\mathbf{T}'$ to be in the $xz$ plane, with $\theta$ being the relative angle.

\section{Results}
\label{NR}

In this section, we present the numerical results of our work. For the physical constants, we take the reduced Planck constant times the speed of light $\hbar c=197.327\ \text{MeV}\!\cdot\!\text{fm}$, the average nucleon mass $m_N=938.918$ MeV, and the fine structure constant $\alpha=1/137.036$. For the effective nucleon-nucleon interaction, we take the Minnesota force \cite{Thompson:1977zz}, with the admixture parameter $u$ taken to be $u=0.94687$ \cite{Theeten:2007zz}. The energy of the free $\alpha$ particle is found to be $E_\alpha=-24.2834$ MeV, with the oscillator parameter being $b=1.36$ fm. The $2\alpha$ threshold energy is then given by $E_\text{th}=2E_\alpha=-48.5668$ MeV.

\subsection{Brink Wave Function \emph{versus} Brink-THSR wave function}
\label{BWFvsBTHSRWF}

\begin{figure}

\centering

\begin{subfigure}[b]{\textwidth}
\centering
\minipage{0.5\textwidth}
  \includegraphics[width=\linewidth]{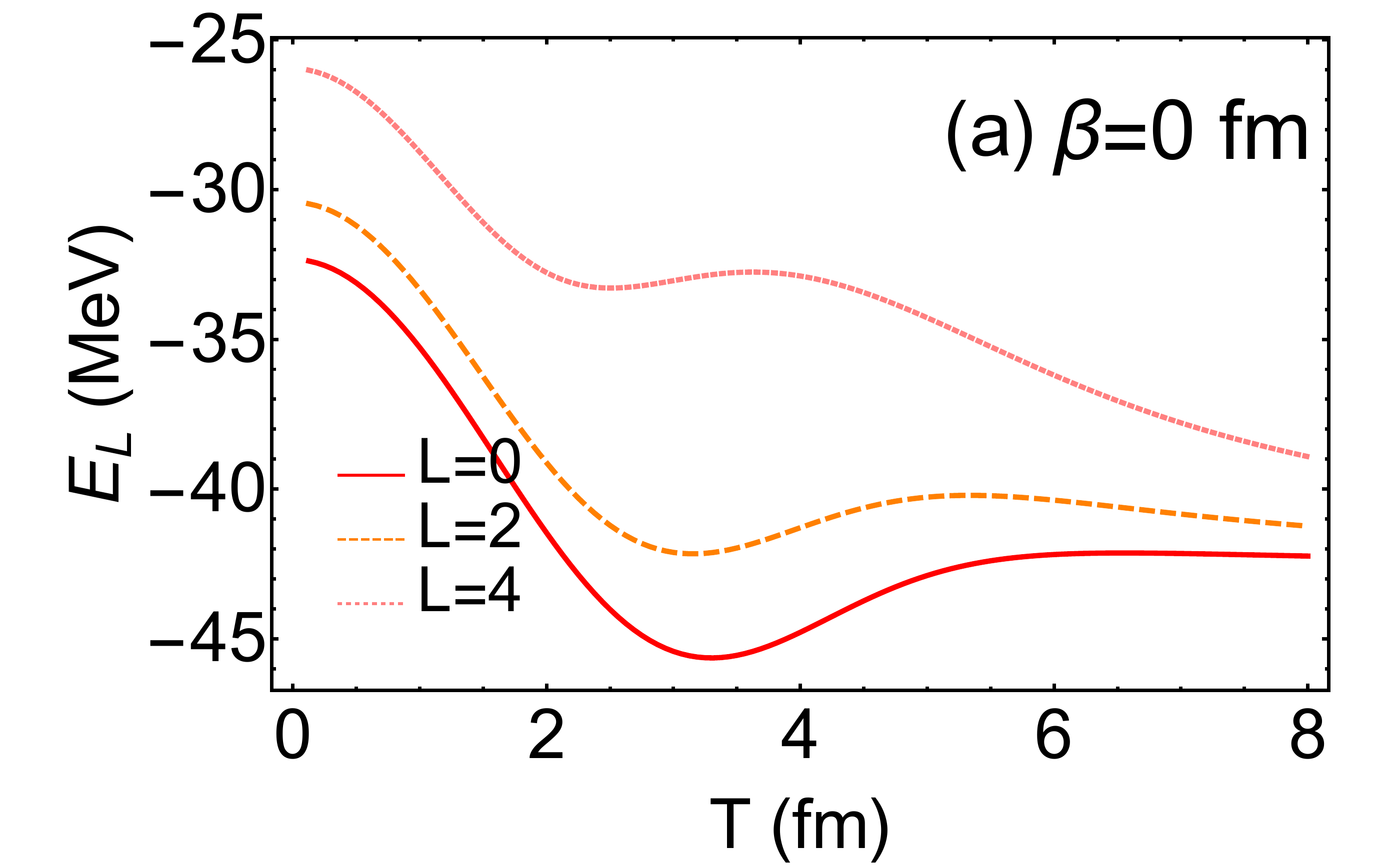}
%  \caption*{$\quad\ \ \ $(a)}
\endminipage\hfill
\minipage{0.5\textwidth}
  \includegraphics[width=\linewidth]{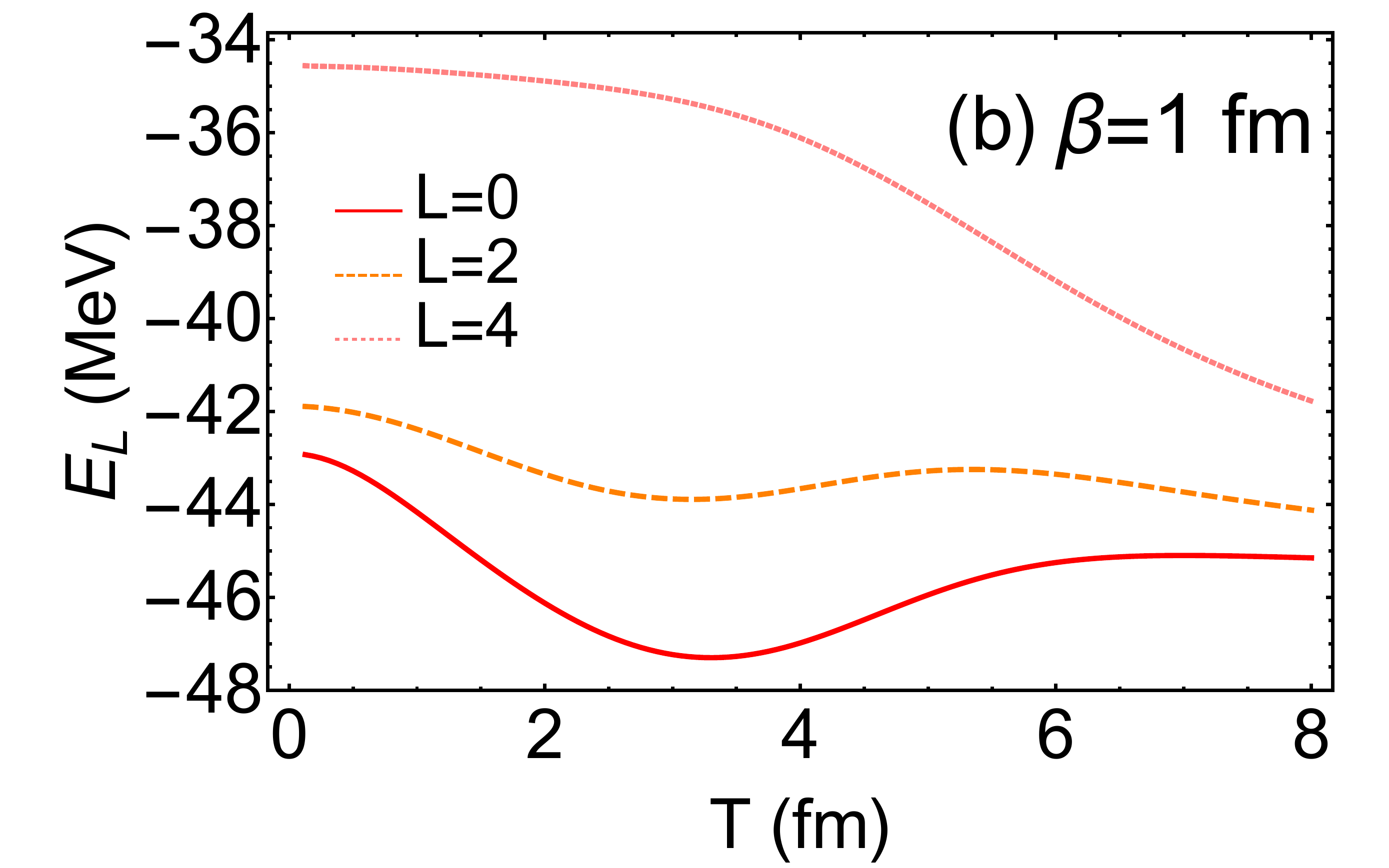}
%  \caption*{$\quad\ \ \ $(b)}
\endminipage
\end{subfigure}

\begin{subfigure}[b]{\textwidth}
\centering
\minipage{0.5\textwidth}
  \includegraphics[width=\linewidth]{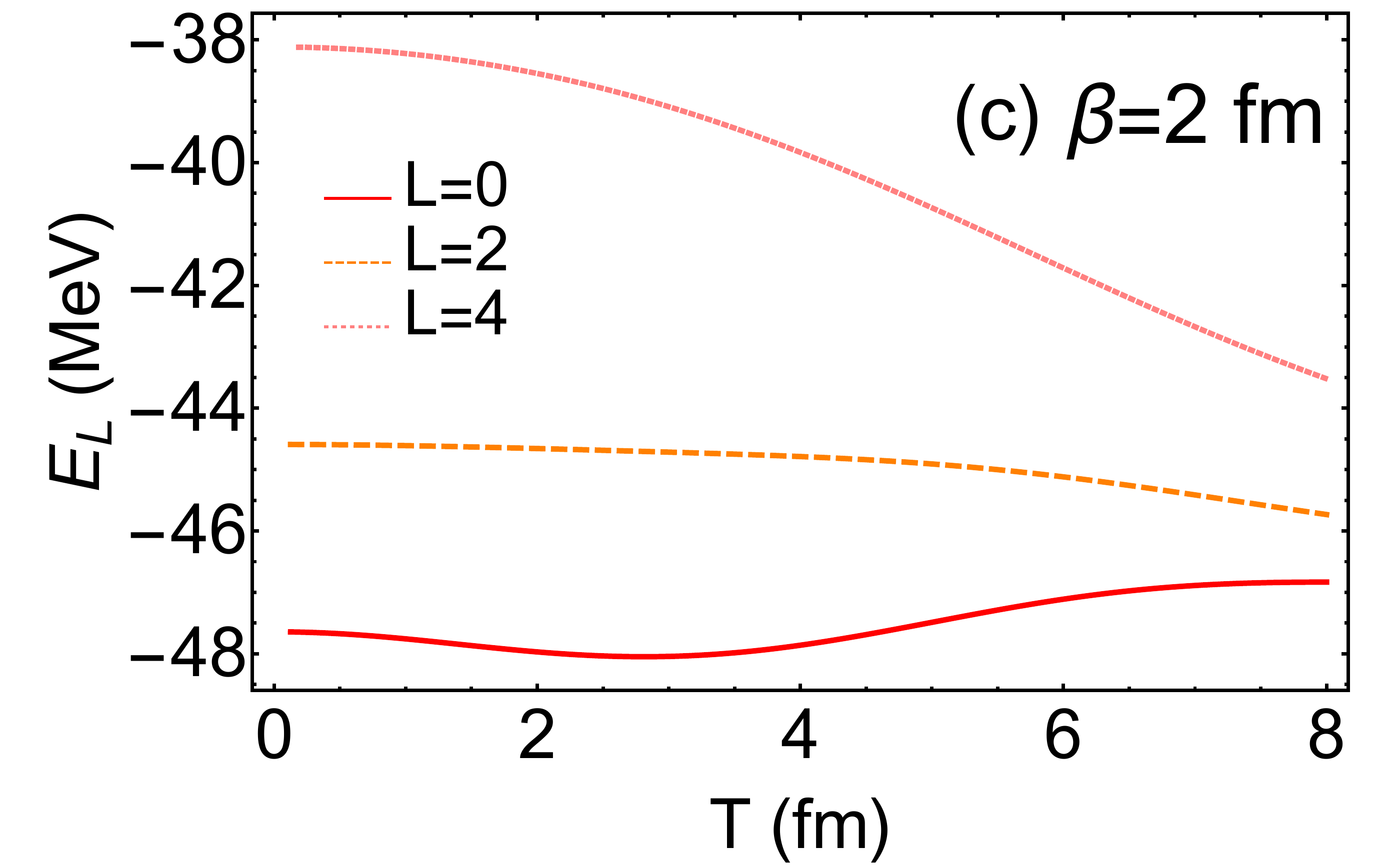}
%  \caption*{$\quad\ \ \ $(c)}
\endminipage\hfill
\minipage{0.5\textwidth}
  \includegraphics[width=\linewidth]{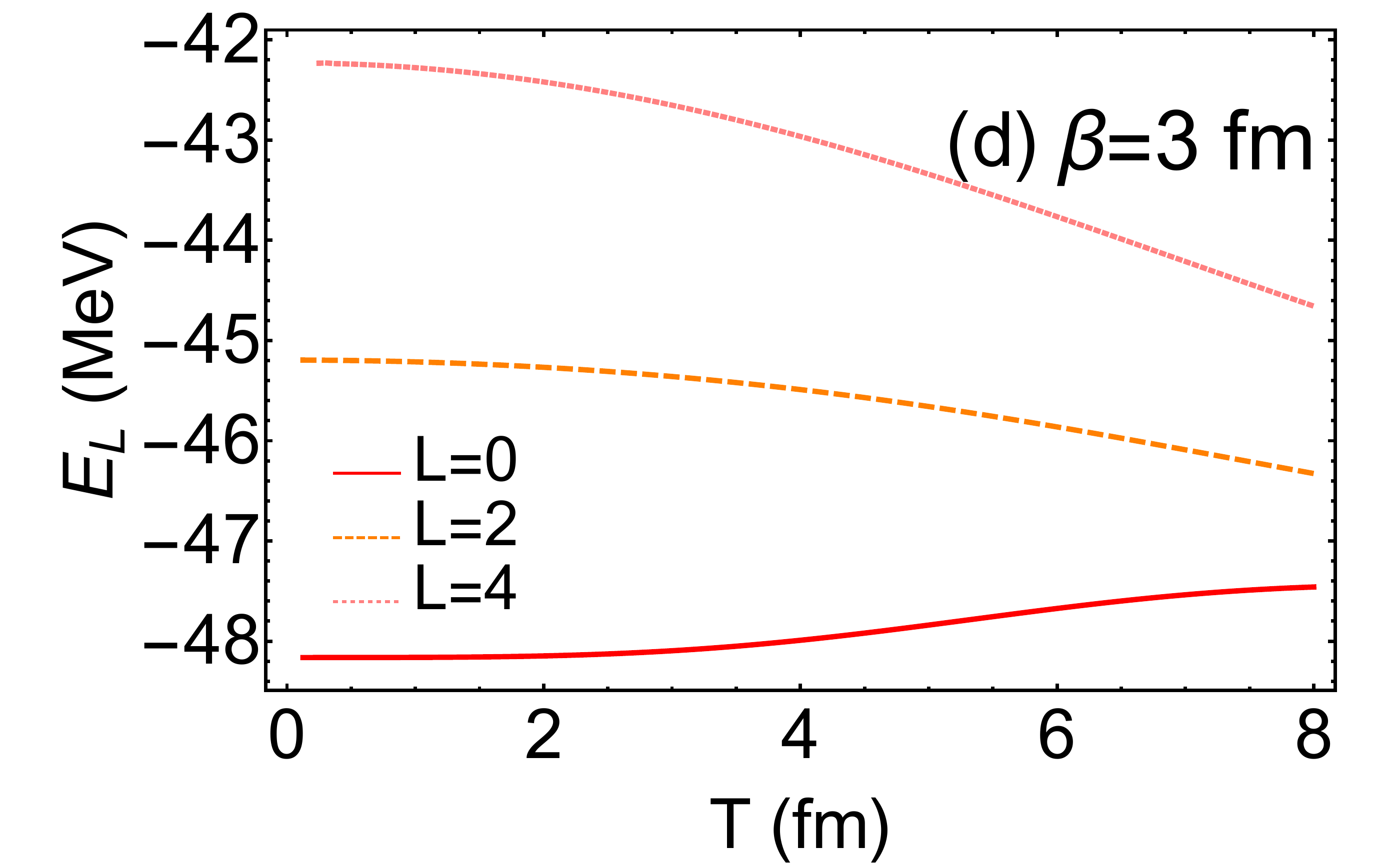}
%  \caption*{$\quad\ \ \ $(d)}
  \endminipage
\end{subfigure}

\caption{The energy curves for the $0^+$, $2^+$, and $4^+$ states of ${}^{8}$Be given by a single Brink-THSR wave function, with the parameter $\beta$ being 0 fm, 1 fm, 2 fm, and 3 fm. For $\beta=0$ fm, the Brink-THSR wave function is reduced to the Brink wave function.}
\label{EC}
\end{figure}

\begin{figure}

\centering

\begin{subfigure}[b]{\textwidth}
\centering
\minipage{0.5\textwidth}
  \includegraphics[width=\linewidth]{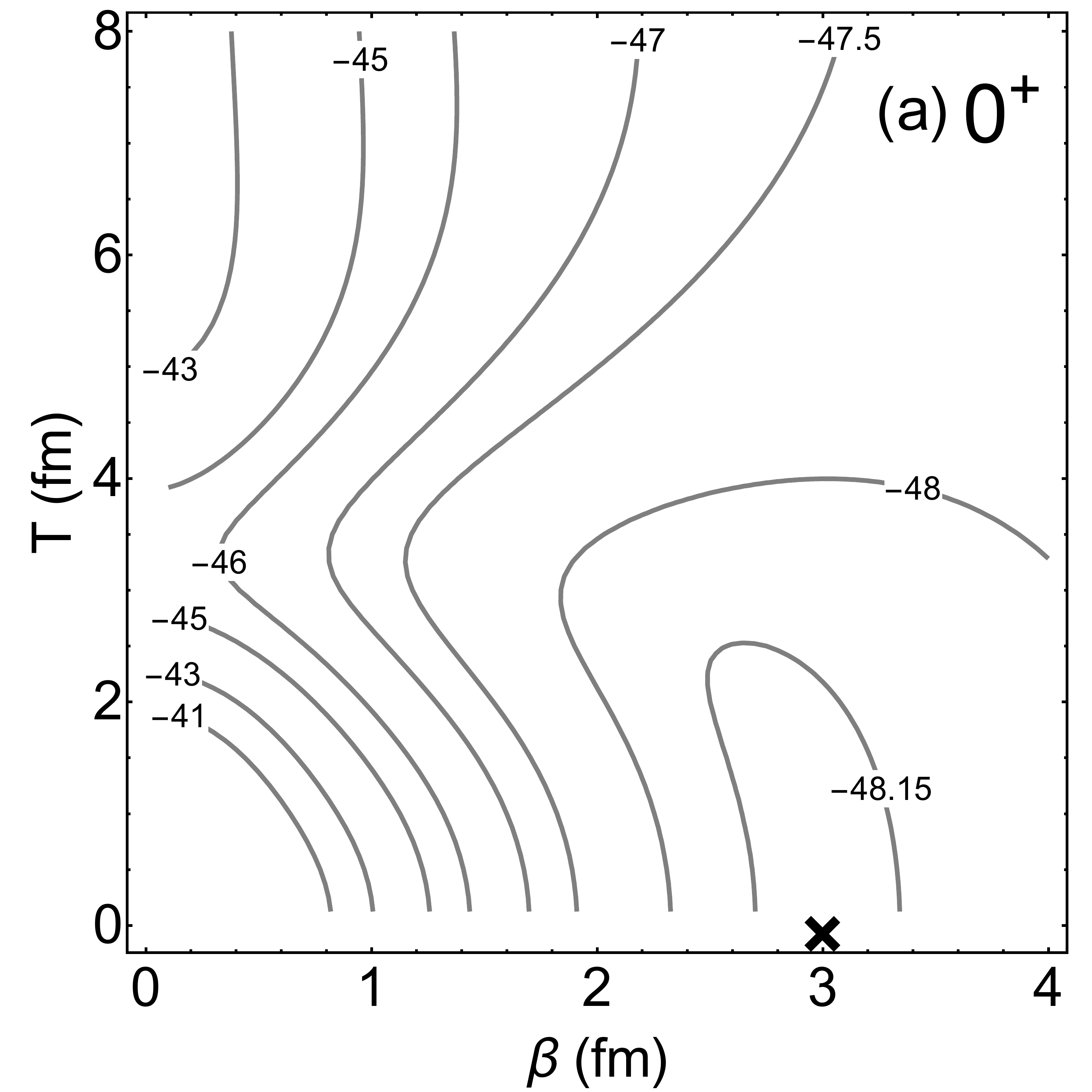}
%  \caption*{$\quad\ \ \ $(a)}
\endminipage\hfill
\minipage{0.5\textwidth}
  \includegraphics[width=\linewidth]{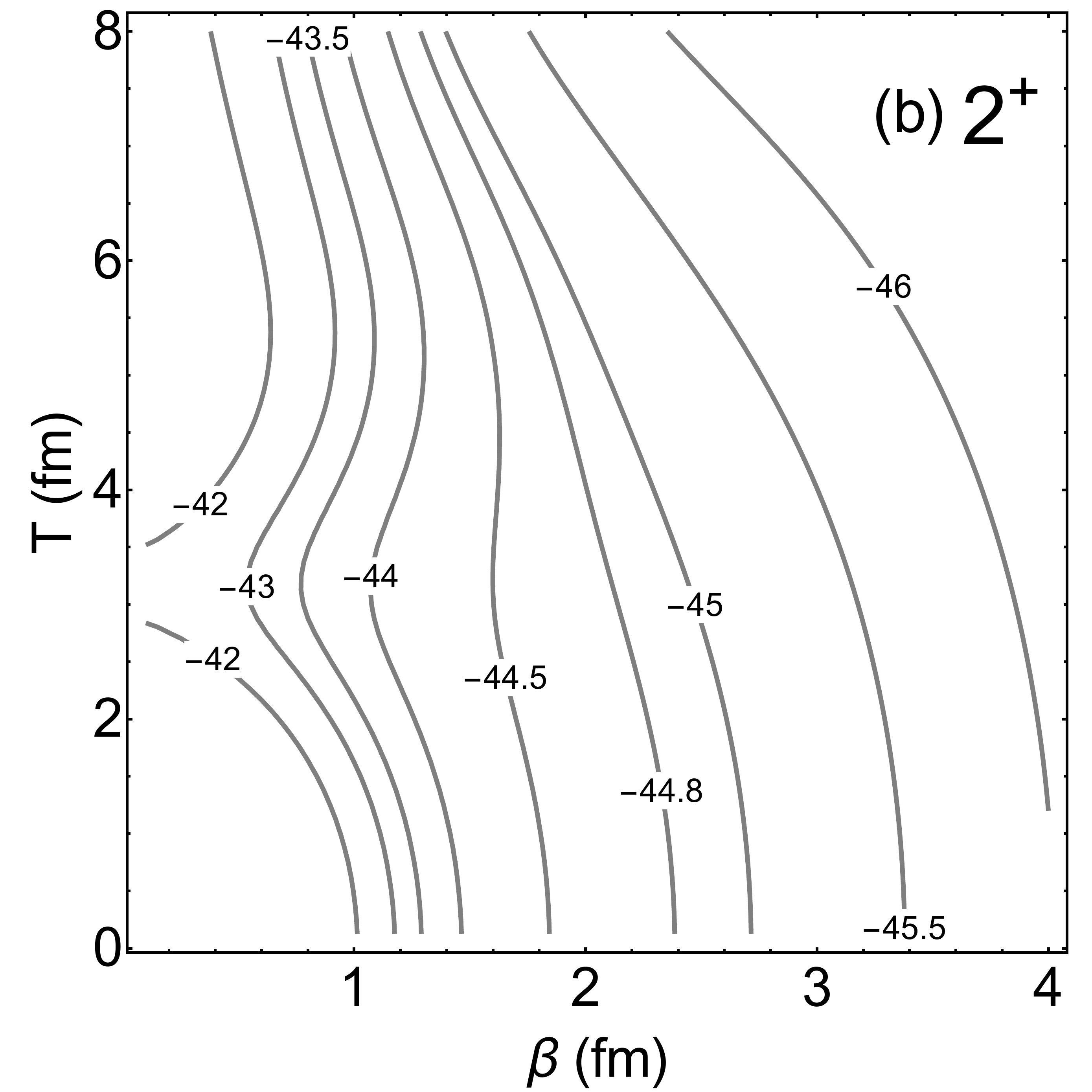}
%  \caption*{$\quad\ \ \ $(b)}
\endminipage
\end{subfigure}

\begin{subfigure}[b]{\textwidth}
\centering
\minipage{0.5\textwidth}
  \includegraphics[width=\linewidth]{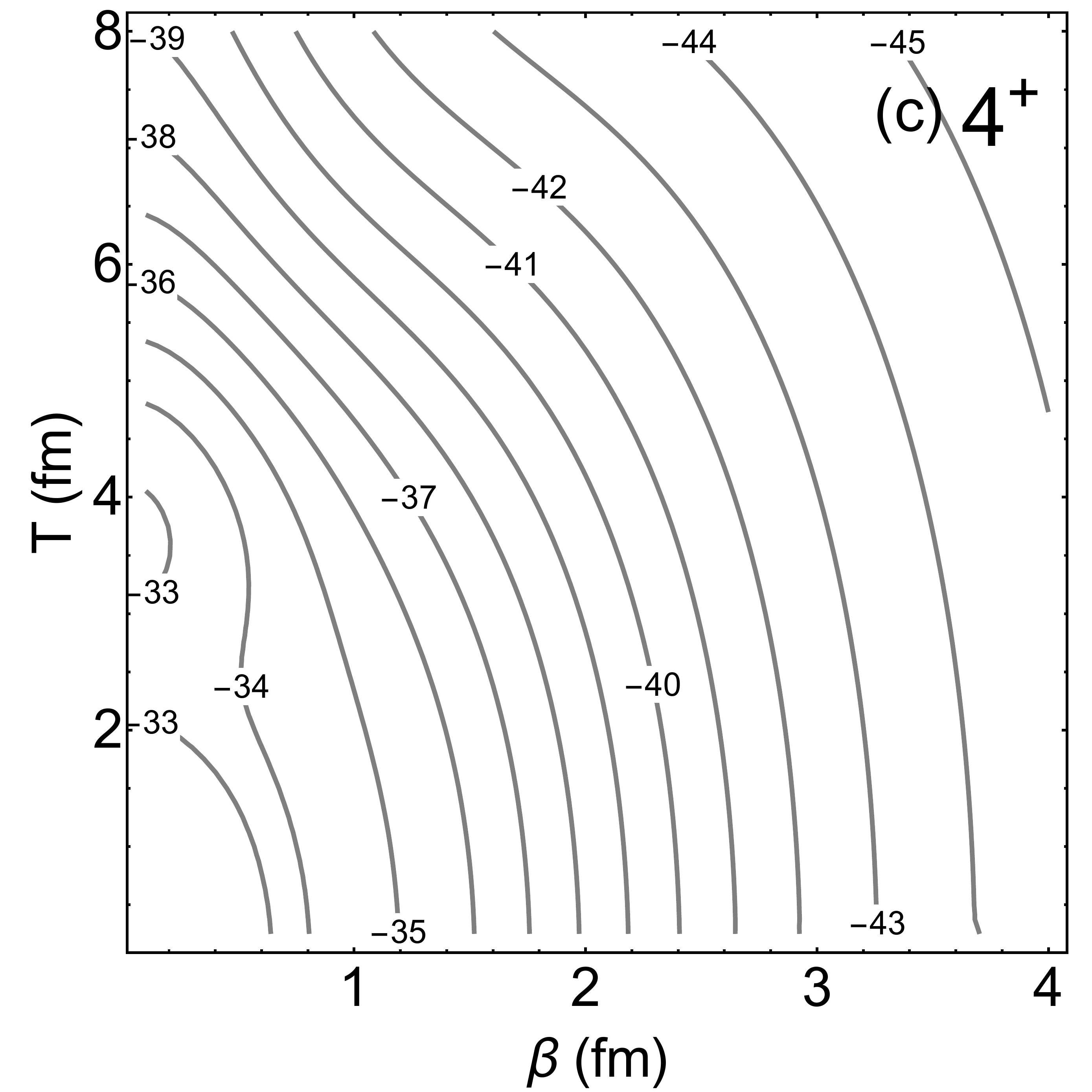}
%  \caption*{$\quad\ \ \ $(c)}
\endminipage
\end{subfigure}

\caption{The energy surfaces for the $0^+$, $2^+$, and $4^+$ states of ${}^{8}$Be given by a single Brink-THSR wave function. The contour labels are the corresponding energies for the $\alpha+\alpha$ system in the unit of MeV. In Fig.~\ref{ES}a, the black cross denotes the local minimum of the energy surface at $(\beta,T)=(3\text{ fm},0\text{ fm})$.}
\label{ES}
\end{figure}

First, we study the low-lying states of the $\alpha+\alpha$ system with a single Brink wave function and a single Brink-THSR wave function in the bound-state approximation. The total energy is given by
\begin{align}
E_{L}(\beta,T)=\frac{\braket{\widehat{\Psi}_{L}(\beta,T)|H_L|\widehat{\Psi}_{L}(\beta,T)}}{\braket{\widehat{\Psi}_{L}(\beta,T)|\widehat{\Psi}_{L}(\beta,T)}}.
\label{EL}
\end{align}
Experimentally, ${}^{8}$Be is found to have three low-lying resonant states, i.e., the $0^+$ state as the ground state with a resonant energy $\mathcal{E}^\text{exp}_{0^+}=0.0918$ MeV above the $2\alpha$ disintegration threshold and a tiny decay width $\Gamma^\text{exp}_{0^+}=5.57$ eV, the $2^+$ state as the first excitation state with a resonant energy $\mathcal{E}^{\text{exp}}_{2^+}=3.12$ MeV and a large decay width $\Gamma^\text{exp}_{2^+}=1.513$ MeV, and the $4^+$ state as the second excitation state with a resonant energy $\mathcal{E}^\text{exp}_{4^+}=11.44$ MeV and a large decay width $\Gamma^\text{exp}_{4^+}\approx3.5$ MeV \cite{Tilley:2004zz} (see also Table \ref{Results4Resonances}).
The numerical results could be found in Fig.~\ref{EC} and \ref{ES}.
In Fig.~\ref{EC}, the energy curves of the $0^+$ (red solid line), $2^+$ (blue dashed line), and $4^+$ (green dotted line) states of the $\alpha+\alpha$ system are plotted, with the parameter $\beta$ being chosen representatively to be $\beta=0$ fm, $1$ fm, $2$ fm, $3$ fm. For $\beta=0$ fm, the Brink-THSR wave function is reduced to the Brink wave function, with the two $\alpha$ clusters being fixed at two endpoints of a ``dumbbell''. The local minima could be found for all the $0^+$, $2^+$, and $4^+$ states, with their energies (corresponding $T$ values) given by $-45.6397$ MeV (3.30 fm), $-42.1630$ MeV (3.15 fm), and $-33.2870$ MeV (2.51 fm), respectively. At the first sight, the results given by a single Brink wave function look good. The obtained excitation energies for the $2^+$ and $4^+$ states are $\mathcal{E}_{2^+}=3.4767$ MeV and $\mathcal{E}_{4^+}=12.3527$ MeV, lying close to the experimental values. However, a careful examination of these results reveals the following shortcomings. The $2^+$ and $4^+$ states are found to sit in the local minima protected by the Coulomb barriers. Usually, this leads to the expectation that the $2^+$ and $4^+$ states are long-lived resonant states with small decay widths. This conflicts with the experimental data, which show that these decay widths are actually quite large. Also, the ground state is found to have a resonant energy of $2.9271$ MeV above the $2\alpha$ disintegration threshold, which is significantly larger than the experimental value of 0.0918 MeV. These shortcomings provide us with important motivations to use the single Brink-THSR wave function with the nonzero $\beta$ parameter to improve the results.
For $\beta=1$ fm, the local minimum of the $4^+$ state disappears. The local minima (the corresponding $T$ values) for the $0^+$ and $2^+$ states persist and are found to be $-47.2982$ MeV (3.30 fm) and $-43.8908$ MeV (3.15 fm). 
For $\beta=2$ fm, the local minima of both the $2^+$ and $4^+$ states disappear, and the local minimum (the corresponding T value) of the $0^+$ state is found to be $-48.0482$ MeV (2.81 fm). 
For $\beta=3$ fm, the local minimum (the corresponding T value) of the $0^+$ state is found to be $-48.1635$ MeV (0 fm). 
One can see that, as the parameter $\beta$ grows from zero, the local minimum of the $0^+$ state persists and is protected by the Coulomb barrier all along. On the contrary, the local minima of the $2^+$ and $4^+$ states disappear successively. These facts suggest that, the $0^+$ state is truly a long-lived resonant state, while the $2^+$ and $4^+$ states are not. These are consistent with the experimental data. Also, the energy of the $0^+$ state decreases continuously as $\beta$ increases from 0 to 3 fm. The resonant energy at $\beta=3$ fm is found to be $\mathcal{E}_{0^+}=0.4033$ MeV, closer to the experimental value.
%For the $2^+$ curves, it is also noticed that, as the energy of the local minimum gets lower, the height of the energy barrier that protects the quasi-stability of the $2^+$ state diminishes accordingly and eventually vanishes when $\beta>2$ fm. Therefore, unlike the $0^+$ state whose local minimum is always protected by a non-vanishing energy barrier, the $2^+$ state of ${}^{8}$Be is expected to have a large decay width. This is confirmed by the experimental data which gives the decay width of the $2^+$ state to be $\Gamma_{2^+}=1.5$ MeV, much larger than the decay width of the $0^+$ state being $\Gamma_{0^+}=5.57$ eV. The realistic asymptotic behavior of the resonant state is needed to get an improved description of the $2^+$ state. Similar results hold also for the $4^+$ state.
%The parameter $\beta$ is found to be about 3 fm for the $0^+$ state, about 1 fm for the $2^+$ state, and about $0$ fm for the $4^+$ state.
In Fig.~\ref{ES}, the energy surfaces of the $0^+$, $2^+$, and $4^+$ states of the $\alpha+\alpha$ system are plotted, giving us another opportunity to better understand the situation.
%, which provide extra valuable information on the physical properties of the $\alpha+\alpha$ system. 
For the $0^+$ state, a local minimum is found at $(\beta,T)=(3\text{ fm},\,0\text{ fm})$, with the corresponding energy being $-48.1635$ MeV, and the Brink-THSR wave function is reduced to the THSR wave function. The $2^+$ and $4^+$ states display different features, and no local minima are found on the $T$-$\beta$ plane. This is consistent with the fact that these two states have large decay widths and the bound-state approximation works less well. The absence of local minima for the $2^+$ and $4^+$ states is also found by Ref.~\cite{Funaki:2002fn}, where the deformed THSR wave function is used in the calculations.
% However, the energy spectrum of the $2^+$ and $4^+$ states might still be extracted approximately by determining the global minimum values of all the possible local minima of the energy curve $E_{l}(\beta,T)$ with a given parameter $\beta$, which gives $E_{2^+}\approx$ 4 MeV and $E_{4^+}\approx$ 14.5 MeV with respect to the $\alpha+\alpha$ threshold energy $E_\text{th}$.
Therefore, the local minima given by the single Brink wave function are actually unstable in the $\beta$ direction, and the $\alpha+\alpha$ system could reduce its energy further by allowing the $\alpha$ clusters to move freely around the endpoints. For the $0^+$ state, the two nuclear containers at the endpoints coalesce to form a big nuclear container. For the $2^+$ and $4^+$ states, the nuclear containers grow up endlessly with no obstructions from the Coulomb barriers and the $\alpha$ clusters move apart to the infinity in the end. 
Last but not least, we would like to mention that, under the antisymmetrization, the functional spaces of the THSR wave function $\widehat{\Psi}_L(\beta,0\text{ fm})$ and the Brink-THSR wave function $\widehat{\Psi}_L(\beta,T\to0\text{ fm})$ are different. The former describes only the spherical $0^+$ states, while the latter describes not only the spherical $0^+$ state but also the non-spherical $2^+$ and $4^+$ states with non-zero spins.
%the Brink-THSR wave function is convenient to study cluster states with non-zero spins. In the limit of $T\to0$, the Brink-THSR wave function in Eq.~(11) is reduced to the spherically symmetric THSR wave function, and 
%it is straightforward to see that $\widehat{\Psi}_L(\beta,0)=0$ for $l\geq2$. Therefore, if it is the spherically symmetric THSR wave function that is started with, the $2^+$ and $4^+$ states can never be reached. On the contrary, the Brink-THSR wave function with the non-zero $T$ parameter is not spherically symmetric. It contains naturally the partial-wave components with non-zero angular momenta and can be used conveniently to study cluster states with non-zero spins. Besides, one could choose to start first with the Brink-THSR wave function with the non-zero $T$ parameter, do all the calculations, and take the limit of $T\to0$ fm in the end. 
Indeed, in the limit of $T\to0$ fm, although $\braket{\widehat{\Psi}_L(\beta,T)|H_L|\widehat{\Psi}_L(\beta,T)}$ and $\braket{\widehat{\Psi}_L(\beta,T)|\widehat{\Psi}_L(\beta,T)}$ become zero for $L\geq2$, their quotient $E_L(\beta,T)=\braket{\widehat{\Psi}_L(\beta,T)|H_L|\widehat{\Psi}_L(\beta,T)}/\braket{\widehat{\Psi}_L(\beta,T)|\widehat{\Psi}_L(\beta,T)}$ is finite and corresponds to the physical observable. 
%Fig.~\ref{EC} shows explicitly that all the $0^+$, $2^+$ and $4^+$ energy curves given by the Brink-THSR wave functions can be extended smoothly to $T=0$ fm without any problems. The discussions here are useful complements to Ref.~\cite{Funaki:2002fn}.
%, where the deformed THSR wave function is proposed to describe the cluster states with non-zero spins.

In this subsection, we compare the energy spectrum of the $\alpha+\alpha$ system given by a single Brink wave function with that given by a single Brink-THSR wave function in the bound-state approximation. It is found that, the Brink-THSR wave function generally gives better theoretical results both qualitatively and quantitatively, identifying correctly the non-quasi-stability of the $2^+$ and $4^+$ states and the quasi-stability of the $0^+$ ground state.  
Very recently, Refs.~\cite{Tohsaki:2017ifu,Tohsaki:2018fmj,Tohsaki:2018srg} suggest the existence of many exotic quasi-stable $\alpha$-cluster structures, such as the fullerene-shaped $\alpha$-cluster structure and the long $\alpha$-chains. In these studies, a single Brink wave function is adopted to model the system. It would be interesting to redo the analysis with the Brink-THSR wave function, which frees the $\alpha$ clusters from their fixed positions and brings another opportunity to better understand these exotic $\alpha$-cluster structures.
%Also, it is worth mentioning that, compared with the spherically symmetric THSR wave function, the Brink-THSR wave function is more convenient to describe the $2^+$ and $4^+$ states. If one starts directly with the spherically symmetric THSR wave function and does not employ any ``mediators'' like the Brink-THSR wave function, generally only the $0^+$ state can be described.
%These show the advantage and the potential of the Brink-THSR wave function.

\subsection{Phase Shifts of the $\alpha+\alpha$ Elastic Scattering}
\label{PSAAES}

\begin{figure}
\centering
\includegraphics[width=0.85\textwidth]{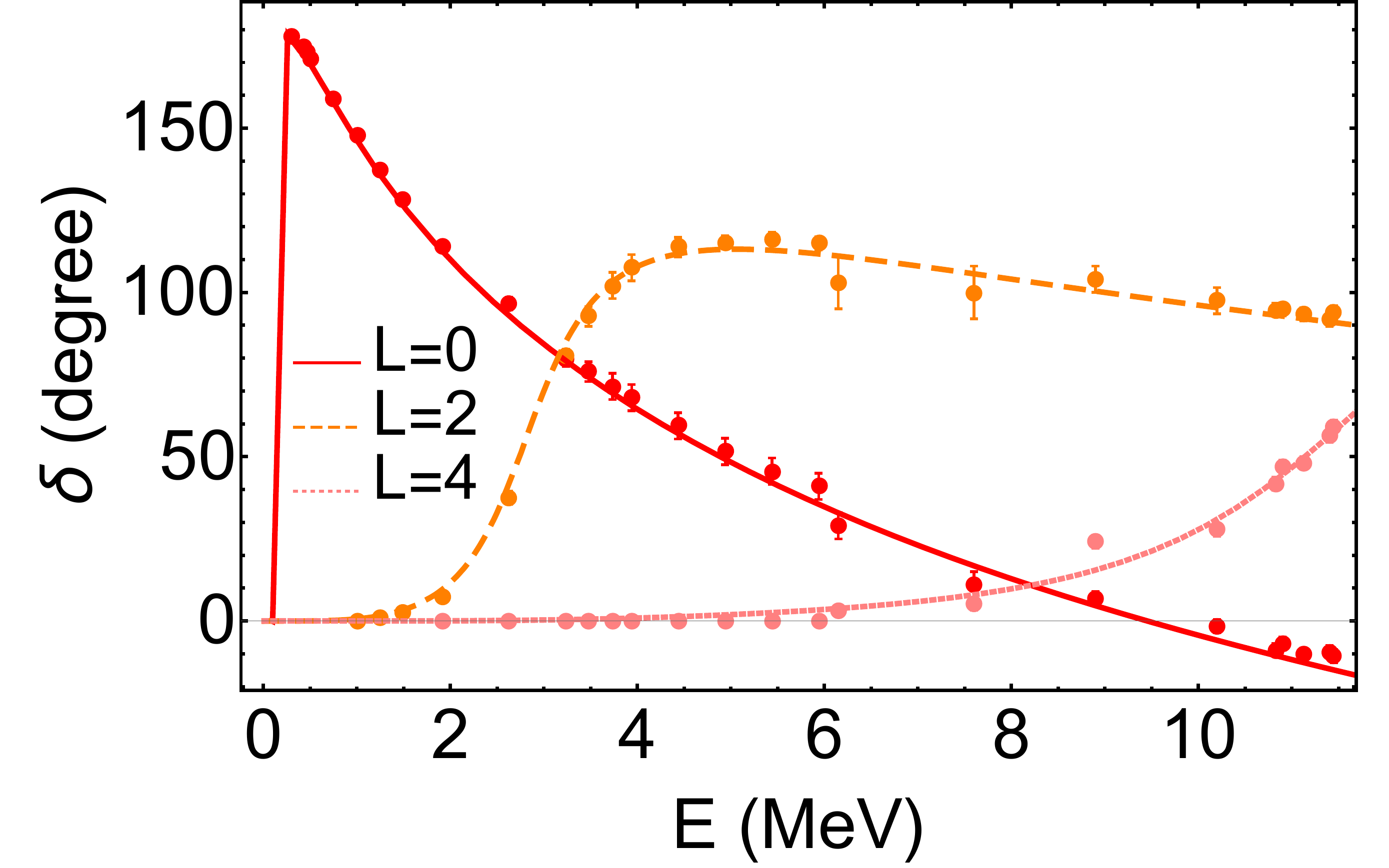}
\caption{The phase shifts for the $\alpha+\alpha$ elastic scattering in the $S$, $D$, and $G$ waves against the total energy of the $\alpha+\alpha$ system in the CM frame. The parameter $\beta$ takes the values of 0 fm, 0.5 fm, and 1 fm. The channel radius is given by $a=7.0$ fm. The numerical results are close to each other for these three cases, and the corresponding plots cannot be distinguished clearly. The data points are experimental data taken from Ref.~\cite{AFZAL:1969zz}.}
\label{PhaseShifts}
\end{figure}

Before carrying out the calculations, we should choose the values of all the auxiliary parameters, including the channel radius $a$ that separates the interior and exterior regions, the discretized generator coordinates $\{T_n\}$ that locate the two nuclear containers at different positions, and the parameter $\beta$ that determines the size of the nuclear containers. Practically, different values of the auxiliary parameters give slightly different numerical results due to the limited model space and the finite working precision. The channel radius $a$ should be chosen large enough but cannot be overlarge. The nuclear interactions and the antisymmetrization effect between two $\alpha$ clusters should be safely neglected in the exterior region, while the exponentially decaying Brink-THSR bases do not become too small at the channel radius to be superposed within the finite copies to match the oscillating exterior wave functions. In this subsection, we choose the channel radius $a\sim7.0$ fm. The generator coordinates $\{T_n\}$ in Eq.~\eqref{IWF} should also be chosen with care. If they are numerically close to each other, the adjacent Hamiltonian and overlap matrix elements could be numerically close, which may cause problems in solving the generalized eigenvalue problem. Also, $\{T_n\}$ cannot be chosen to be overlarge, since this will result in almost vanishing matrix elements that may cause troubles for the generalized eigenvalue solver as well. As a benchmark, we take $\{T_n\}$ from 0.8 fm to 8 fm in step of 0.8 fm. The parameter $\beta$ takes three different values $\beta=0$ fm, $0.5$ fm, and $1$ fm. 

The numerical results for the phase shifts could be found in Fig.~\ref{PhaseShifts} with the channel radius $a=7.0$ fm.
The theoretical results given by these three $\beta$ values turn out to be numerically close to each other, and cannot be distinguished clearly in Fig.~\ref{PhaseShifts}. The experimental data are taken from Ref.~\cite{AFZAL:1969zz} and plotted as data points. It is straightforward to see that the theoretical results agree well with the experimental data.
In Table \ref{Table4BetaDependence}, some representative values of the phase shifts given by $\beta=0$ fm, $0.5$ fm, and $1$ fm are shown explicitly. For $\beta=0$ fm, the Brink-THSR wave function is reduced to the Brink wave function, and our method corresponds to the GCM $+$ the $R$-matrix theory (a.k.a.~the microscopic $R$-matrix theory in Refs.~\cite{Baye:1977vpg,Descouvemont:2010cx,Descouvemont:2012,Baye:1974dkx,Baye:1992zz}) and thus is mathematically equivalent to the RGM.
% In other words, our calculations at $\beta=0$ fm correspond to the GCM and RGM calculations.
From Table \ref{Table4BetaDependence} one can see that, the results given by $\beta=0.5$ fm and $1$ fm are numerically consistent with those given by $\beta=0$ fm (i.e., GCM/RGM). 
To check the consistency of our formalism, we also study the channel-radius dependence of the phase shifts. Some representative results are given in Table \ref{Table4ChannelRadiusDependence} with the parameters $\beta=1$ fm and $\{T_n\}=\{0.8\text{ fm}, 1.6\text{ fm}, \cdots, 7.2\text{ fm}, 8.0\text{ fm}\}$. In the case of $(a,l)=(8.0\text{ fm}, 4)$, we adopt $\{T_n\}=\{0.8\text{ fm}, 1.6\text{ fm}, \cdots, 7.2\text{ fm}, 8.0\text{ fm}, 8.8 \text{ fm}\}$ to make the results convergent. For different channel radii at $a=7.0\text{ fm}, 8.0\text{ fm}, 9.0\text{ fm}$, the partial-wave phase shifts at the same reaction energies agree with each other numerically within few percents. This is good enough for our current purposes. 
%There are also few noticeable exceptions at, e.g., $(a, l)=(8.0\text{ fm}, 4)$, which are actually related to the small model space and finite working precision adopted in the calculations. Better agreements can be obtained with a larger set of $\{T_n\}=\{0.8\text{ fm}, 1.6\text{ fm}, \cdots, 8.8\text{ fm}\}$, which gives the $G$-wave phase shifts at $E=1 \text{ MeV}, 5 \text{ MeV}, 10 \text{ MeV}, 15 \text{ MeV}$ to be $\delta=0.0022^{\circ}, 1.9498^\circ, 27.9296^\circ, 119.6904^\circ$, respectively, certainly agreeing better with other results at $a=7.0 \text{ fm}$ and $9.0 \text{ fm}$.

\begin{table}
\caption{The $\beta$ dependence of the phase shifts with different reaction energies and partial waves. The phase shifts are given in the unit of degree. In this table, we take $a=7.0$ fm and $\{T_n\}=\{0.8\text{ fm}, 1.6\text{ fm}, \cdots, 7.2\text{ fm}, 8.0\text{ fm}\}$. The case of $\beta=0$ fm corresponds to the GCM and thus is equivalent to the RGM.}
\label{Table4BetaDependence}
%\begin{center}
\centering
\begin{tabular}{cccc}
\hline
\hline
& \multicolumn{2}{c}{{\centering $S$-Wave Elastic Scattering with $l=0$} }\\
\hline
\hspace{2.5mm}$E$ $(\text{MeV})$\hspace{2.5mm} & \hspace{2.5mm}$\beta=0$ fm (GCM/RGM) \hspace{2.5mm} & \hspace{2.5mm}$\beta=0.5$ fm\hspace{2.5mm} & \hspace{2.5mm}$\beta=1$ fm\hspace{2.5mm}\\
\hline
\hspace{2.5mm}$1$\hspace{2.5mm} & \hspace{2.5mm}$146.0603$\hspace{2.5mm} & \hspace{2.5mm}$146.1205$\hspace{2.5mm} &  \hspace{2.5mm}$146.1788$\hspace{2.5mm} \\
\hspace{2.5mm}$5$\hspace{2.5mm} &  \hspace{2.5mm}$48.3671$\hspace{2.5mm} & \hspace{2.5mm}$48.3464$\hspace{2.5mm} &  \hspace{2.5mm}$48.3171$\hspace{2.5mm} \\
\hspace{2.5mm}$10$\hspace{2.5mm} &  \hspace{2.5mm}$-4.8431$\hspace{2.5mm} &  \hspace{2.5mm}$-4.5708$\hspace{2.5mm} &  \hspace{2.5mm}$-4.3510$\hspace{2.5mm} \\
\hspace{2.5mm}$15$\hspace{2.5mm} &  \hspace{2.5mm}$-36.8503$\hspace{2.5mm} & \hspace{2.5mm}$-36.7729$\hspace{2.5mm} &  \hspace{2.5mm}$-36.7085$\hspace{2.5mm} \\
\hline
\hline
      & \multicolumn{2}{c}{{\centering $D$-Wave Elastic Scattering with $l=2$} }\\
\hline
\hspace{2.5mm}$E$ $(\text{MeV})$\hspace{2.5mm} &  \hspace{2.5mm}$\beta=0$ fm (GCM/RGM) \hspace{2.5mm} &  \hspace{2.5mm}$\beta=0.5$ fm\hspace{2.5mm} &  \hspace{2.5mm}$\beta=1$ fm\hspace{2.5mm}\\
\hline
\hspace{2.5mm}$1$\hspace{2.5mm} &  \hspace{2.5mm}$0.6000$\hspace{2.5mm} & \hspace{2.5mm}$0.6187$\hspace{2.5mm} &  \hspace{2.5mm}$0.6379$\hspace{2.5mm} \\
\hspace{2.5mm}$5$\hspace{2.5mm} & \hspace{2.5mm}$112.8403$\hspace{2.5mm} & \hspace{2.5mm}$112.9675$\hspace{2.5mm} &  \hspace{2.5mm}$113.1474$\hspace{2.5mm} \\
\hspace{2.5mm}$10$\hspace{2.5mm} & \hspace{2.5mm}$95.9473$\hspace{2.5mm} &  \hspace{2.5mm}$96.0567$\hspace{2.5mm} &  \hspace{2.5mm}$96.1221$\hspace{2.5mm} \\
\hspace{2.5mm}$15$\hspace{2.5mm} &  \hspace{2.5mm}$78.9347$\hspace{2.5mm} & \hspace{2.5mm}$79.1715$\hspace{2.5mm} &  \hspace{2.5mm}$79.3553$\hspace{2.5mm} \\
\hline
\hline
      & \multicolumn{2}{c}{{\centering $G$-Wave Elastic Scattering with $l=4$} }\\
\hline
\hspace{2.5mm}$E$ $(\text{MeV})$\hspace{2.5mm} &  \hspace{2.5mm}$\beta=0$ fm (GCM/RGM) \hspace{2.5mm} & \hspace{2.5mm}$\beta=0.5$ fm\hspace{2.5mm} & \hspace{2.5mm}$\beta=1$ fm\hspace{2.5mm}\\
\hline
\hspace{2.5mm}$1$\hspace{2.5mm} & \hspace{2.5mm}$0.0005$\hspace{2.5mm} & \hspace{2.5mm}$0.0007$\hspace{2.5mm} &  \hspace{2.5mm}$0.0015$\hspace{2.5mm} \\
\hspace{2.5mm}$5$\hspace{2.5mm} &  \hspace{2.5mm}$1.3944$\hspace{2.5mm} &  \hspace{2.5mm}$1.5326$\hspace{2.5mm} &  \hspace{2.5mm}$1.9335$\hspace{2.5mm} \\
\hspace{2.5mm}$10$\hspace{2.5mm} & \hspace{2.5mm}$27.6251$\hspace{2.5mm} &  \hspace{2.5mm}$27.6609$\hspace{2.5mm} &  \hspace{2.5mm}$27.7600$\hspace{2.5mm}\\
\hspace{2.5mm}$15$\hspace{2.5mm} &  \hspace{2.5mm}$119.7914$\hspace{2.5mm} & \hspace{2.5mm}$119.8328$\hspace{2.5mm} &  \hspace{2.5mm}$119.4784$\hspace{2.5mm} \\
\hline
\hline
\end{tabular}

%\end{center}
\end{table}

\begin{table}
\caption{The channel-radius dependence of the phase shifts with different reaction energies and partial waves. The phase shifts are given in the unit of degree. In this table, we take $\beta=1$ fm and $\{T_n\}=\{0.8\text{ fm}, 1.6\text{ fm}, \cdots, 7.2\text{ fm}, 8.0\text{ fm}\}$, except the case of $(a,l)=(8.0\text{ fm}, 4)$, where we adopt $\{T_n\}=\{0.8\text{ fm}, 1.6\text{ fm}, \cdots, 7.2\text{ fm}, 8.0\text{ fm}, 8.8 \text{ fm}\}$ to converge the results.}
\label{Table4ChannelRadiusDependence}
\centering
%\begin{center}
\begin{tabular}{cccc}
\hline
\hline
      & \multicolumn{2}{c}{{\centering $S$-Wave Elastic Scattering with $l=0$} }\\
\hline
\hspace{2.5mm}$E$ $(\text{MeV})$\hspace{2.5mm} & \hspace{2.5mm}$a=7.0$ fm\hspace{2.5mm} & \hspace{2.5mm}$a=8.0$ fm\hspace{2.5mm} & \hspace{2.5mm}$a=9.0$ fm\hspace{2.5mm}\\
\hline
\hspace{2.5mm}$1$\hspace{2.5mm} & \hspace{2.5mm}$146.1788$\hspace{2.5mm} & \hspace{2.5mm}$146.2826$\hspace{2.5mm} &  \hspace{2.5mm}$146.1025$\hspace{2.5mm} \\
\hspace{2.5mm}$5$\hspace{2.5mm} &  \hspace{2.5mm}$48.3171$\hspace{2.5mm} & \hspace{2.5mm}$48.3982$\hspace{2.5mm} &  \hspace{2.5mm}$48.3248$\hspace{2.5mm} \\
\hspace{2.5mm}$10$\hspace{2.5mm} &  \hspace{2.5mm}$-4.3510$\hspace{2.5mm} &  \hspace{2.5mm}$-4.0317$\hspace{2.5mm} &  \hspace{2.5mm}$-4.0909$\hspace{2.5mm} \\
\hspace{2.5mm}$15$\hspace{2.5mm} &  \hspace{2.5mm}$-36.7085$\hspace{2.5mm} & \hspace{2.5mm}$-36.5629$\hspace{2.5mm} &  \hspace{2.5mm}$-36.7323$\hspace{2.5mm} \\
\hline
\hline
      & \multicolumn{2}{c}{{\centering $D$-Wave Elastic Scattering with $l=2$} }\\
\hline
\hspace{2.5mm}$E$ $(\text{MeV})$\hspace{2.5mm} &  \hspace{2.5mm}$a=7.0$ fm\hspace{2.5mm} &  \hspace{2.5mm}$a=8.0$ fm\hspace{2.5mm} &  \hspace{2.5mm}$a=9.0$ fm\hspace{2.5mm}\\
\hline
\hspace{2.5mm}$1$\hspace{2.5mm} &  \hspace{2.5mm}$0.6379$\hspace{2.5mm} & \hspace{2.5mm}$0.6893$\hspace{2.5mm} &  \hspace{2.5mm}$0.5671$\hspace{2.5mm} \\
\hspace{2.5mm}$5$\hspace{2.5mm} & \hspace{2.5mm}$113.1474$\hspace{2.5mm} & \hspace{2.5mm}$113.1185$\hspace{2.5mm} &  \hspace{2.5mm}$113.0804$\hspace{2.5mm} \\
\hspace{2.5mm}$10$\hspace{2.5mm} & \hspace{2.5mm}$96.1221$\hspace{2.5mm} &  \hspace{2.5mm}$96.7556$\hspace{2.5mm} &  \hspace{2.5mm}$96.5777$\hspace{2.5mm} \\
\hspace{2.5mm}$15$\hspace{2.5mm} &  \hspace{2.5mm}$79.3553$\hspace{2.5mm} & \hspace{2.5mm}$79.2905$\hspace{2.5mm} &  \hspace{2.5mm}$79.0443$\hspace{2.5mm} \\
\hline
\hline
      & \multicolumn{2}{c}{{\centering $G$-Wave Elastic Scattering with $l=4$} }\\
\hline
\hspace{2.5mm}$E$ $(\text{MeV})$\hspace{2.5mm} &  \hspace{2.5mm}$a=7.0$ fm\hspace{2.5mm} & \hspace{2.5mm}$a=8.0$ fm\hspace{2.5mm} & \hspace{2.5mm}$a=9.0$ fm\hspace{2.5mm}\\
\hline
\hspace{2.5mm}$1$\hspace{2.5mm} & \hspace{2.5mm}$0.0015$\hspace{2.5mm} & \hspace{2.5mm}$0.0022$\hspace{2.5mm} &  \hspace{2.5mm}$0.0013$\hspace{2.5mm} \\
\hspace{2.5mm}$5$\hspace{2.5mm} &  \hspace{2.5mm}$1.9335$\hspace{2.5mm} &  \hspace{2.5mm}$1.9498$\hspace{2.5mm} &  \hspace{2.5mm}$1.9254$\hspace{2.5mm} \\
\hspace{2.5mm}$10$\hspace{2.5mm} & \hspace{2.5mm}$27.7600$\hspace{2.5mm} &  \hspace{2.5mm}$27.9296$\hspace{2.5mm} &  \hspace{2.5mm}$27.8978$\hspace{2.5mm}\\
\hspace{2.5mm}$15$\hspace{2.5mm} &  \hspace{2.5mm}$119.4784$\hspace{2.5mm} & \hspace{2.5mm}$119.6904$\hspace{2.5mm} &  \hspace{2.5mm}$120.0626$\hspace{2.5mm} \\
\hline
\hline
\end{tabular}

%\end{center}
\end{table}

\begin{figure}
\centering
\includegraphics[width=0.75\textwidth]{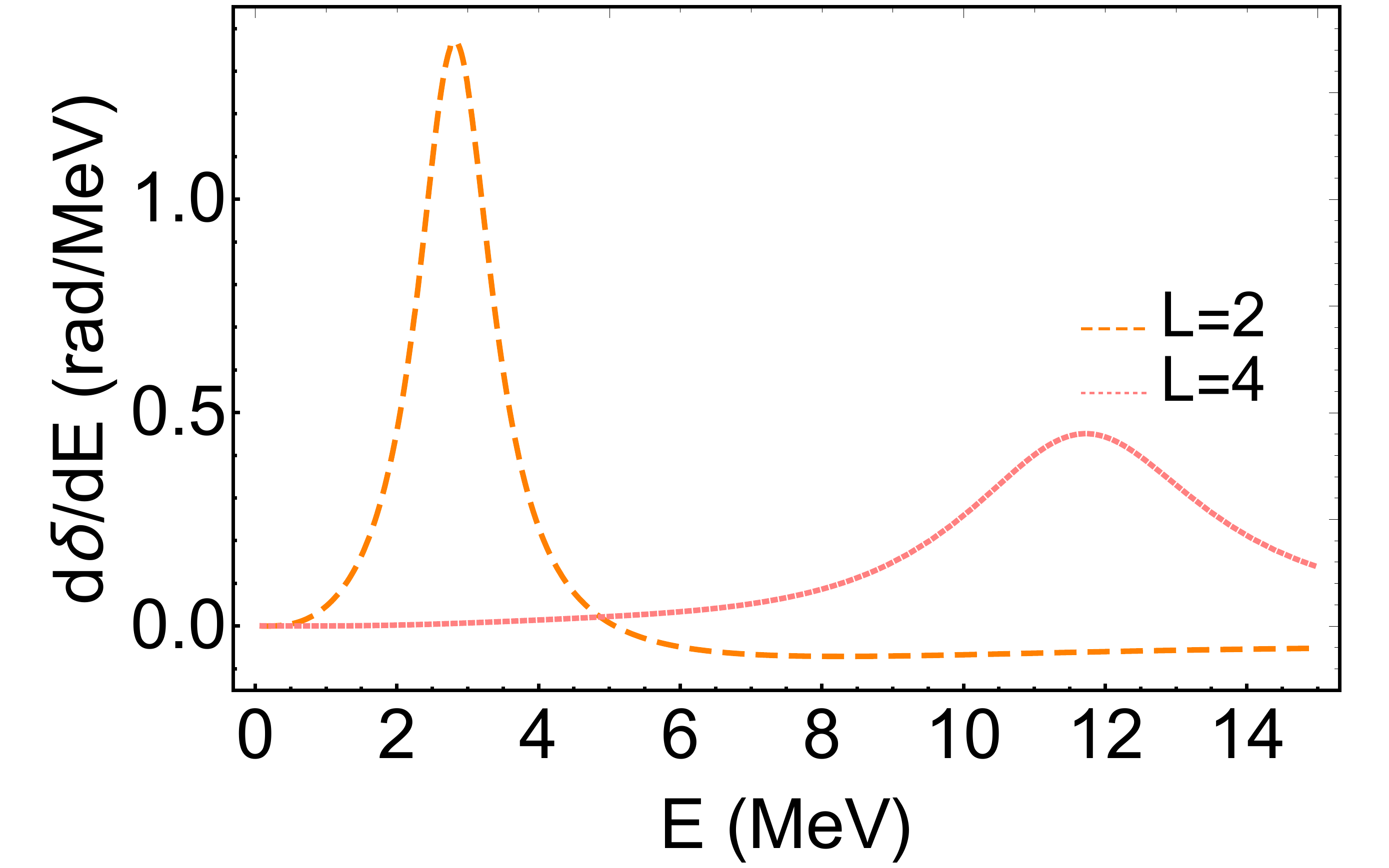}
\caption{$\mathrm{d}\delta/\mathrm{d}E$ (in the unit of radian/MeV) of the $\alpha+\alpha$ elastic scattering in the $D$ and $G$ waves against the total energy of the $\alpha+\alpha$ system in the CM frame.}
\label{DeltavsE}
\end{figure}

%From Fig.~\ref{PhaseShifts}, one could determine further the energies and the decay widths of the $2^+$ and $4^+$ states, which are absent in previous studies on the non-localized cluster model. In the pure Breit-Wigner approximation with the vanishing background phase shift \cite{Thompson:2009}, the phase shift in the vicinity of the resonance takes the form of
%\begin{align}
%\delta(E)=\arctan\left(\frac{\Gamma_\text{res}/2}{E_\text{res}-E}\right)+n(E)\pi,\label{BWA}
%\end{align}
%where $E_\text{res}$ and $\Gamma_\text{res}$ are the energy and the decay width of the resonance, and $n(E)$ is the energy-dependent integer added to make $\delta(E)$ a smooth function. Eq.~\eqref{BWA} gives that $\delta(E_\text{res})=90^\circ$ and $\delta(E_\text{res}-\Gamma_\text{res}/2)=45^\circ$, which can be used to determine the energy and the decay width of the target resonance. The energy (the decay width) of the $2^+$ state with respect to the $2\alpha$ disintegration threshold $E_\text{th}$ is found to be $E_{2^+}=3.34$ MeV $(\Gamma_{2^+}=1.33\text{ MeV})$, which is close to the experimental values $E^\text{exp}_{2^+}=3.030$ MeV $(\Gamma^\text{exp}_{2^+}=1.513 \text{ MeV})$. The energy (the decay width) of the $4^+$ state is found to be $E_{4^+}=12.81$ MeV $(\Gamma_{4^+}=3.70\text{ MeV})$, which is also close to the experimental values $E^\text{exp}_{4^+}=11.350$ MeV $(\Gamma^\text{exp}_{4^+}\approx3.5 \text{ MeV})$. 

Given the numerical results on the phase shifts, one could determine the resonant energies and decay widths of the ${2}^{+}$ and $4^+$ states in the framework of the hermitian quantum mechanics. The energy dependence of the phase shift is given by
\begin{align}
\frac{\mathrm{d}\delta}{\mathrm{d}E}\approx\frac{2\Gamma}{4(E-\mathcal{E}_\text{res})^2+\Gamma^2},
\end{align}
in the Breit-Wigner approximation (see, e.g., Refs.~\cite{Thompson:2009,Moiseyev:2011}) and is plotted in Fig.~\ref{DeltavsE} for the $D$ and $G$ waves. Different from Fig.~\ref{PhaseShifts}, the phase shifts here are in the unit of radian rather than degree. The resonant energy $\mathcal{E}_\text{res}$ is given by the local maximum of $\mathrm{d}\delta/\mathrm{d}E$, while the decay width is given by
\begin{align}
\Gamma\approx2\bigg/\frac{\mathrm{d}\delta}{\mathrm{d}E}\bigg|_{E=\mathcal{E}_\text{res}}.
\end{align}
The numerical results are listed in Table \ref{Results4Resonances}. Unlike the $2^+$ and $4^+$ states, it is not easy to extract accurate information on the $0^+$ state due to its small resonant energy and tiny decay width.

\subsection{Low-Lying Resonant States of ${}^{8}$Be}

\begin{figure}

\centering

\begin{subfigure}[b]{\textwidth}
\centering
\minipage{0.5\textwidth}
  \includegraphics[width=\linewidth]{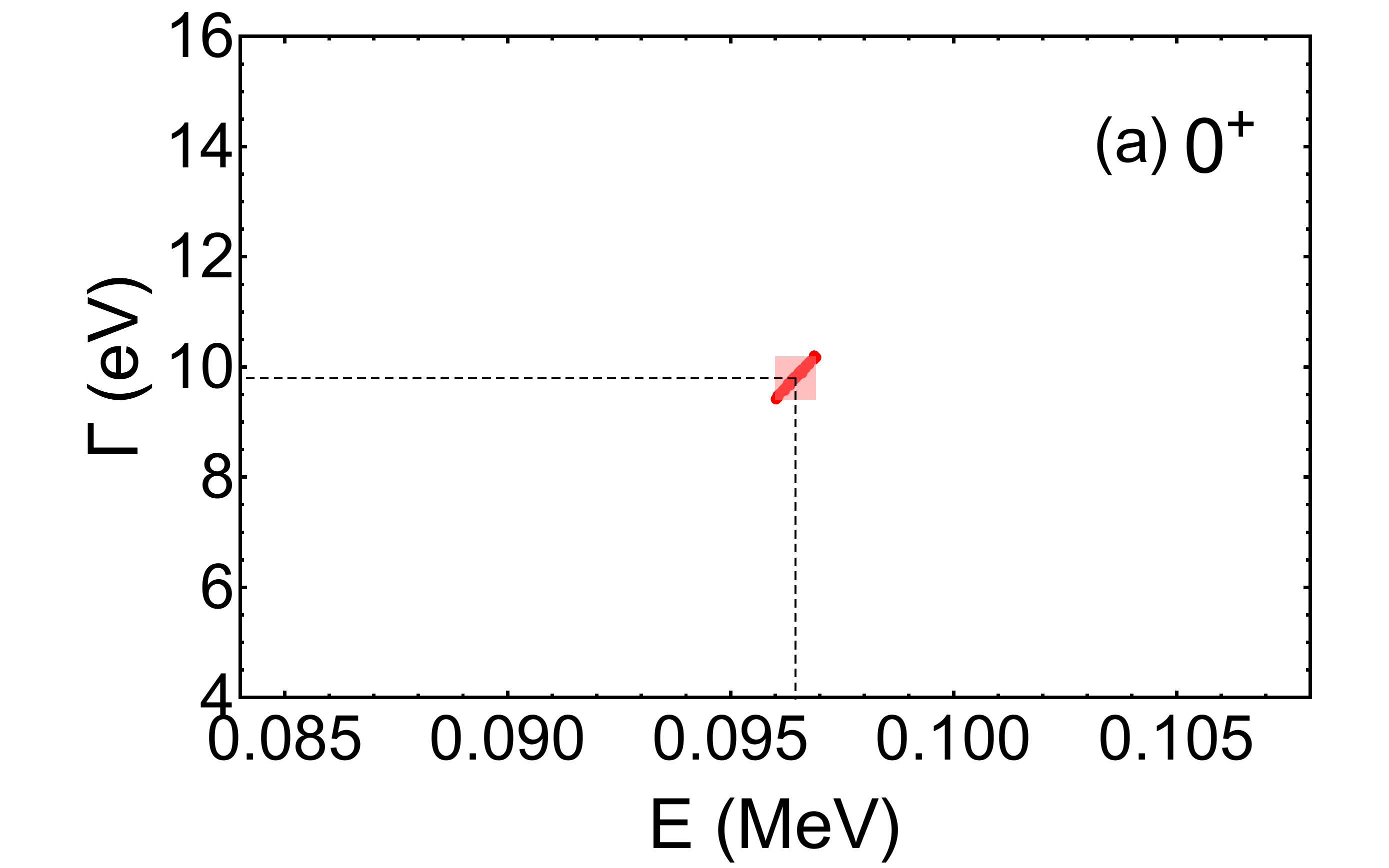}
%  \caption*{$\quad\ \ \ $(a)}
\endminipage\hfill
\minipage{0.5\textwidth}
  \includegraphics[width=\linewidth]{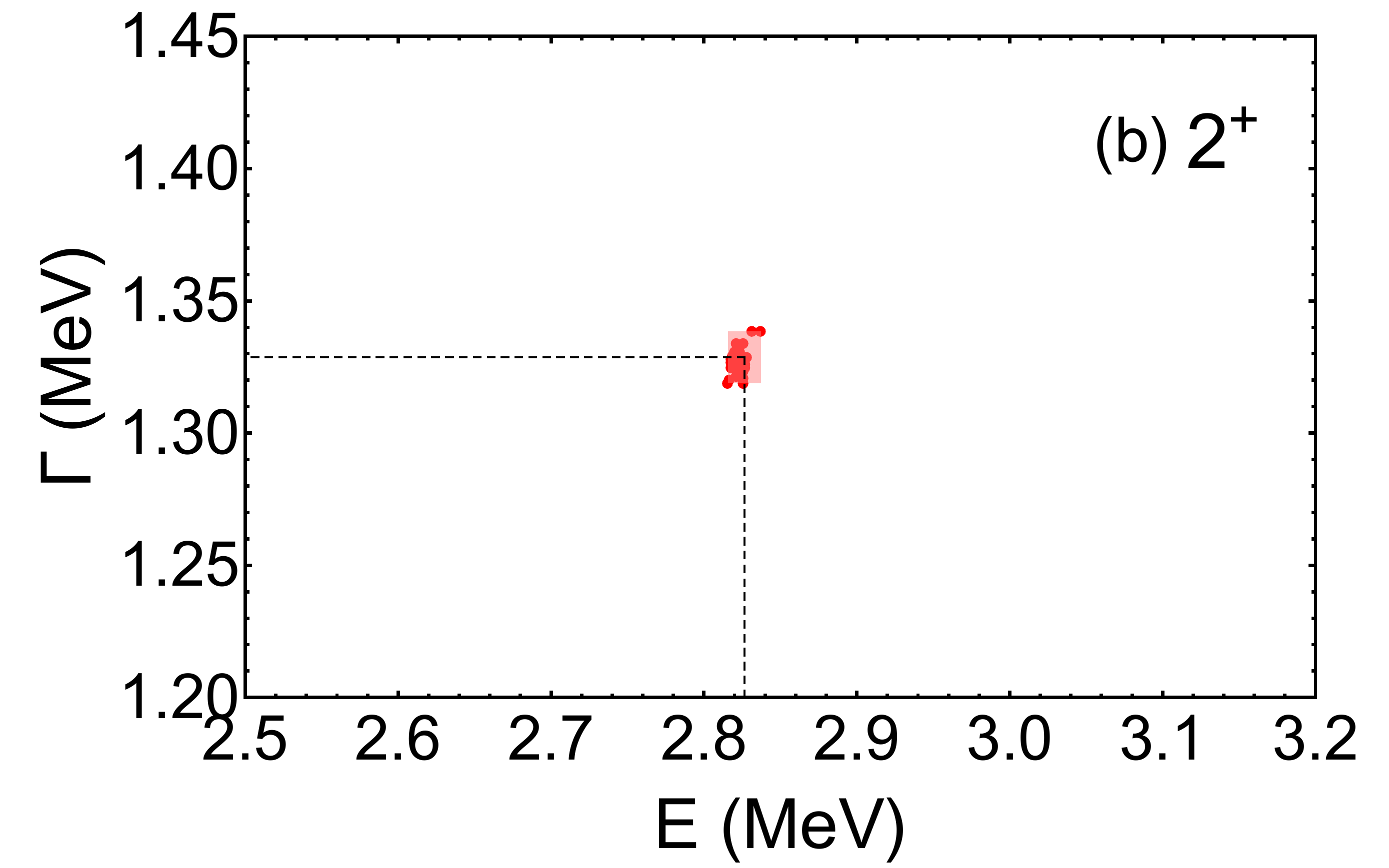}
%  \caption*{$\quad\ \ \ $(b)}
\endminipage
\end{subfigure}

\begin{subfigure}[b]{\textwidth}
\centering
\minipage{0.5\textwidth}
  \includegraphics[width=\linewidth]{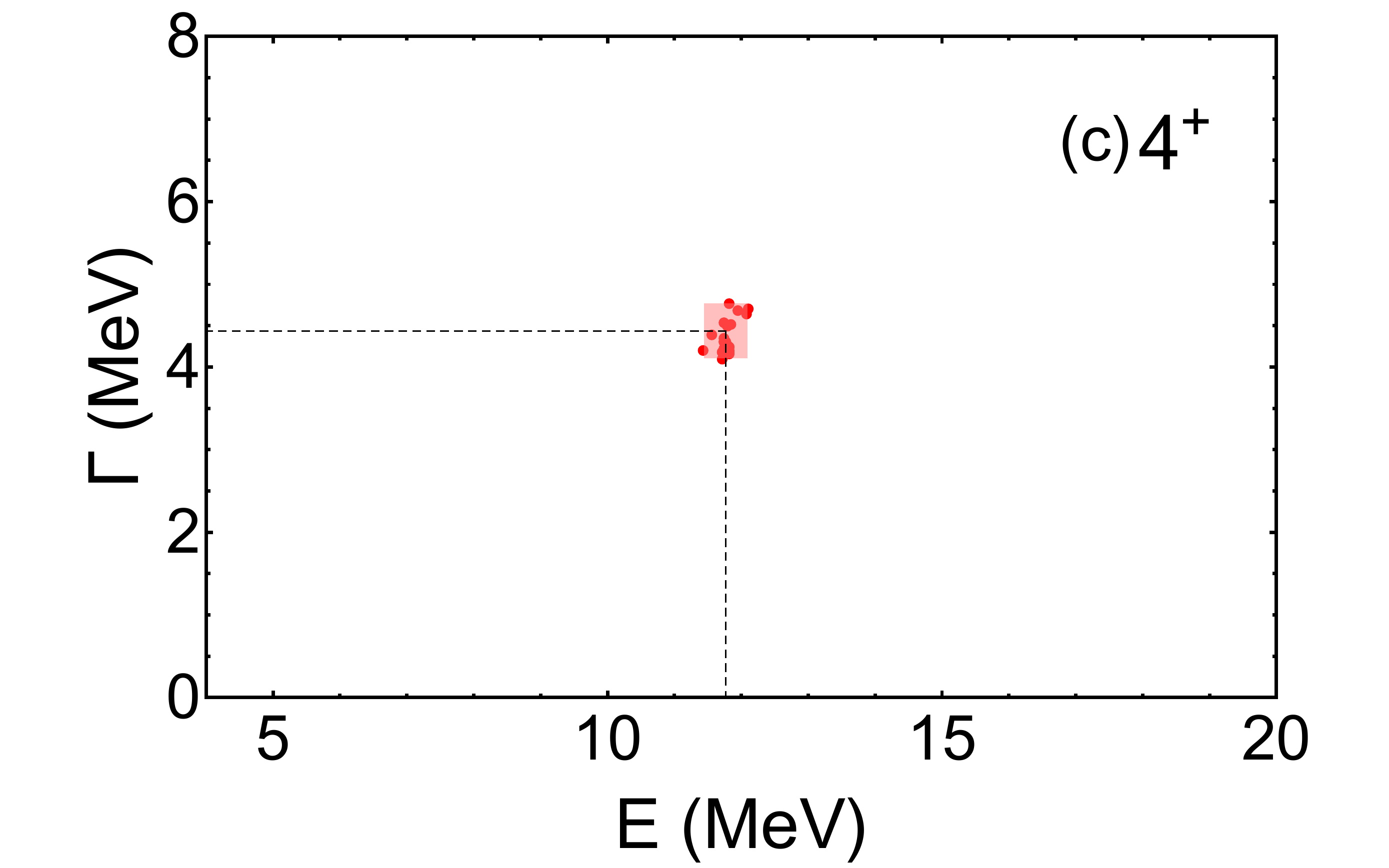}
%  \caption*{$\quad\ \ \ $(c)}
\endminipage
\end{subfigure}

\caption{Resonant energies and decay widths for the low-lying resonances of ${}^{8}$Be from different sets of auxiliary parameters. The data points are the theoretical results. The pink rectangles are the minimal axis-aligned rectangles containing all the data points.}
\label{CE4Res}
\end{figure}

Subsection \ref{BWFvsBTHSRWF} shows that, the realistic asymptotic forms of the resonant states play a significant role in studying the $2^+$ and $4^+$ states, which have large decay widths and cannot be treated consistently in the bound-state approximation. In Subsection \ref{PSAAES}, the resonant energies and decay widths are obtained from the phase-shift data for the $2^+$ and $4^+$ states. In this subsection, we use the formalism developed in Section \ref{TF} to calculate the energy spectrum of ${}^{8}$Be self-consistently from Eq.~\eqref{GEP4RS}, without referring to the $\alpha+\alpha$ elastic scattering process. Some representative iteration processes could be found in Table \ref{Table4Iteration}, where we take $\beta=1$ fm, $a=7.0$ fm. The generator coordinates are given by $\{T_n\}=\{0.1 \text{ fm}, 1.5 \text{ fm}, \cdots, 7.1 \text{ fm}, 8.5 \text{ fm}\}$ for the $0^+$ state, $\{T_n\}=\{0.1 \text{ fm}, 1.667 \text{ fm}, \cdots, 6.333 \text{ fm}, 7.5 \text{ fm}\}$ for the $2^+$ state, and $\{T_n\}=\{0.1 \text{ fm}, 2.2 \text{ fm}, \cdots, 6.4 \text{ fm}, 8.5 \text{ fm}\}$ for the $4^+$  state. Typically, after iterations for $12\sim20$ times, the complex energies get convergent to required precisions. We also try other parameter sets for the auxiliary parameters $(a, \beta, \{T_n\})$. The complex energies are numerically close to each other but the idealized exact agreement cannot be achieved due to the limited model space and finite working precision. In the bound-state calculations, the parameter set giving the lowest energy would be favored by the variational principle. For the resonant-state calculations, no such selection rules are available. There are indeed the complex analog of the variational principle in literature, but it is a stationary principle and cannot be used to put any upper or lower bound on the resonant energy and the decay width \cite{Moiseyev:2011}. Therefore, instead of sticking to a particular parameter set, we have done the calculations using many of them. The resonant energies and decay widths are all plotted in Fig.~\ref{CE4Res}. The spread of the numerical results provides a preliminary estimation of the numerical uncertainties of our calculations. The final results for the resonant energies and decay widths are listed in Table \ref{Results4Resonances}, along with their numerical uncertainties. Good agreement is achieved between the theoretical results and the experimental data. Moreover, we calculate the resonant energy of the $0^+$ state by using the standard GCM in the bound-state approximation. Thanks to its narrow decay width, the bound-state approximation should be applicable. With $\{T_n\}=\{0.1\text{ fm}, 0.845 \text{ fm}, \cdots, 14.255 \text{ fm}, 15 \text{ fm}\}$, the resonant energy is $\mathcal{E}_{0^+}=0.1034$ MeV, quite close to the value given in Table \ref{Results4Resonances}. This could be viewed as another check of the correctness of our calculations. In Subsection \ref{BWFvsBTHSRWF}, the single THSR wave function with $\beta=3$ fm is favored energetically by minimizing the total energy in Eq.~\eqref{EL}. 
%The squared overlap between this single THSR wave function and the GCM wave function in the bound-state approximation is about 0.93. This means that, the GCM wave function could be well expressed by the single THSR wave function. The squared overlap is slightly smaller than that in Refs.~\cite{Funaki:2002fn,Funaki:2009fc}. This might be related partially to the different effective nucleon-nucleon interaction used in this work. Also, those references adopt the deformed THSR wave function to calculate the squared overlap, which gives the improved performance over the spherically symmetric THSR wave function used here.
Refs.~\cite{Funaki:2002fn,Funaki:2009fc} show that, in the bound-state approximation the GCM wave function could be well approximated by the single THSR wave function.
Given the closeness of the resonant energy from our method and the GCM and the narrowness of the decay width of the $0^+$ state, it is reasonable to believe that the real part of the interior wave function from our method shares the same characteristics. 
%A detailed comparison of these wave functions is interesting and important. It lies, however, beyond the scope of this work and is likely to be addressed in the future.  
We do an explicit calculation by taking $\beta=0$ fm for simplicity, where our method is reduced to the GCM + the $R$-matrix theory. We take $a=7$ fm and $\{T_n\}=\{0.1\text{ fm}, 0.9778\text{ fm}, \cdots, 7.1222 \text{ fm}, 8\text{ fm}\}$. The resonant energy is given by $\mathcal{E}_{0^+}=0.0963$ MeV, while the decay width is given by $\Gamma_{0^+}=9.6$ eV. It is found that the squared overlap between the interior wave function and the single THSR wave function with $\beta=3$ fm is about 0.99, which means that the interior wave function of the $0^+$ resonant state is indeed well described by a single THSR wave function, even after taking the realistic boundary condition into consideration. 
%For the more general Brink-THSR wave functions, we expect similar conclusions. While interesting and important, these calculations lie beyond the scope of this work and are likely to be addressed in the future.

\begin{table}
\caption{Iteration solutions of the Bloch-Schr\"odinger equation for the low-lying resonant states of ${}^{8}$Be. The complex energies are given in the unit of MeV. In this table, we take $\beta=1$ fm, $a=7.0$ fm. The generator coordinates are given by $\{T_n\}=\{0.1 \text{ fm}, 1.5 \text{ fm}, \cdots, 7.1 \text{ fm}, 8.5 \text{ fm}\}$ for the $0^+$ state, $\{T_n\}=\{0.1 \text{ fm}, 1.667 \text{ fm}, \cdots, 6.333 \text{ fm}, 7.5 \text{ fm}\}$ for the $2^+$  state, and $\{T_n\}=\{0.1 \text{ fm}, 2.2 \text{ fm}, \cdots, 6.4 \text{ fm}, 8.5 \text{ fm}\}$ for the $4^+$  state.}
\label{Table4Iteration}
\centering
%\begin{center}
\begin{tabular}{cccc}
\hline
\hline
 \hspace{2mm}Iterations \hspace{2mm} & \hspace{6mm}$L=0$ \hspace{6mm} & \hspace{6mm}$L=2$ \hspace{6mm} & \hspace{6mm}$L=4$ \hspace{6mm} \\
\hline
 \hspace{2mm}1 \hspace{2mm} & \hspace{4mm}$0.1$ \hspace{4mm} & \hspace{4mm}$3$ \hspace{4mm} & \hspace{4mm}$20-2.5i$ \hspace{4mm} \\
 \hspace{2mm}2 \hspace{2mm} & \hspace{4mm}$0.09622-8.1197\times10^{-6}i$ \hspace{4mm} & \hspace{4mm}$2.9692 -0.5601 i$ \hspace{4mm} & \hspace{4mm}$11.6042 -1.1359 i$ \hspace{4mm} \\
 \hspace{2mm}3 \hspace{2mm} & \hspace{4mm}$0.09700 -4.1253\times10^{-6}i$ \hspace{4mm} & \hspace{4mm}$2.8748 -0.6641 i$ \hspace{4mm} & \hspace{4mm}$11.3162 -2.2401 i$ \hspace{4mm} \\
 \hspace{4mm}$\cdots$ \hspace{4mm} & \hspace{4mm}$\cdots$ \hspace{4mm} & \hspace{4mm}$\cdots$ \hspace{4mm} & \hspace{4mm}$\cdots$ \hspace{4mm} \\
 \hspace{2mm}12 \hspace{2mm} & \hspace{4mm}$0.09687 -5.0984\times10^{-6} i$ \hspace{4mm} & \hspace{4mm}$2.8190 -0.6636 i$ \hspace{4mm} & \hspace{4mm}$11.8490 -2.2588 i$ \hspace{4mm} \\
 \hspace{2mm}13 \hspace{2mm} & \hspace{4mm}$0.09687 -5.0984\times10^{-6} i$ \hspace{4mm} & \hspace{4mm}$2.8190 -0.6636 i$ \hspace{4mm} & \hspace{4mm}$11.8460 -2.2579 i$ \hspace{4mm} \\
  \hspace{2mm}$14$ \hspace{2mm} & \hspace{4mm}$\cdots$ \hspace{4mm} & \hspace{4mm}$\cdots$ \hspace{4mm} & \hspace{4mm}$11.8459 -2.2599 i$ \hspace{4mm} \\
   \hspace{4mm}$\cdots$ \hspace{4mm} & \hspace{4mm}$\cdots$ \hspace{4mm} & \hspace{4mm}$\cdots$ \hspace{4mm} & \hspace{4mm}$\cdots$ \hspace{4mm} \\
 \hspace{2mm}20 \hspace{2mm} & \hspace{4mm}$\cdots$ \hspace{4mm} & \hspace{4mm}$\cdots$ \hspace{4mm} & \hspace{4mm}$11.8466 -2.2594 i$ \hspace{4mm} \\
 \hspace{2mm}21 \hspace{2mm} & \hspace{4mm}$\cdots$ \hspace{4mm} & \hspace{4mm}$\cdots$ \hspace{4mm} & \hspace{4mm}$11.8466 -2.2594 i$ \hspace{4mm} \\
\hline
\hline
\end{tabular}

%\end{center}
\end{table}

\begin{table}

\caption{Resonant energies and decay widths for the low-lying resonances of ${}^{8}$Be. The experimental data are taken from Ref.~\cite{Tilley:2004zz}. The theoretical values I are given by the phase-shift calculations. The theoretical values II are given by the self-consistent Bloch-Schr\"odinger equations.}
\label{Results4Resonances}
\centering
%\begin{center}
\begin{tabular}{ccccccccc}
\hline
\hline
      & \multicolumn{2}{c}{Experimental Values} 
      & \multicolumn{2}{c}{Theoretical Values I}
      & \multicolumn{2}{c}{Theoretical Values II} \\
      \hline
\hspace{1mm}$L$\hspace{1mm}  
      & \hspace{2mm} $\mathcal{E}$ (MeV) \hspace{4mm} 
      & \hspace{4mm} $\Gamma$ (MeV) \hspace{2mm}
      & \hspace{2mm} $\mathcal{E}$ (MeV) \hspace{4mm} 
      & \hspace{4mm} $\Gamma$ (MeV) \hspace{2mm}
      & \hspace{2mm} $\mathcal{E}$ (MeV) \hspace{4mm} 
      & \hspace{4mm} $\Gamma$ (MeV) \hspace{2mm}
       \\
\hline
0 & 0.0918 & $5.57\times10^{-6}$ & $-$ & $-$ & $0.0965\pm0.0005$ & $(9.8\pm0.4)\times10^{-6}$ \\
2 & 3.12 & 1.513 & 2.82 & 1.46& $2.83\pm0.01$ & $1.33\pm0.01$ \\
4 & 11.44 & 3.5 & 11.73 & 4.44 & $11.8\pm0.3$ & $4.4\pm0.3$ \\
\hline
\hline
\\
\\
\end{tabular}

%\end{center}
\end{table}

\section{Conclusions}
\label{CC}

Recent years witness the proposal and the development of the non-localized cluster model. It provides a new understanding of the nuclear cluster effects based on the picture of non-localized clustering and has been applied to study structural properties of cluster states in various light nuclear systems. In this work, the non-localized cluster model is generalized from bound and quasi-bound states to resonant and scattering states, with the $\alpha+\alpha$ system taken as an example to test the formalism. Following the $R$-matrix theory, the full configuration space is divided into the interior and exterior regions by a channel radius, which has to be chosen properly to make the nuclear interactions and the antisymmetrization effects vanish between different clusters in the exterior region. In the interior region, the Brink-THSR wave functions, the hybrid trial wave functions that combine features of both the Brink and THSR wave functions, are adopted to realize mathematically the picture of non-localized clustering. They are superposed to give the full interior wave functions. The Bloch-Schr\"odinger equation is adopted to match the interior wave functions with the exterior ones given by either the purely outgoing Coulomb-Hankel functions for the resonant states or some combinations of the incoming and outgoing Coulomb-Hankel functions for the scattering states. The single Brink-THSR wave function is adopted to study the low-lying states of ${}^{8}$Be. Compared with the single Brink wave function, the Brink-THSR wave function correctly identifies the non-quasi-stability of the $2^+$ and $4^+$ states and gives the better result on the resonant energy for the $0^+$ state. The phase shifts of the $\alpha+\alpha$ elastic scattering and the properties of the low-lying resonances of $^{8}$Be are studied by solving the Bloch-Schr\"odinger equations with different exterior wave functions. The phase shifts are found to agree well with the experimental data. The Bloch-Schr\"odinger equations for the resonant states are solved self-consistently, and the theoretical values are consistent with those given by the phase-shift calculations, as well as the experimental data.

The study here could be generalized in several directions. First, it is physically important to continue improving the microscopic studies on the $\alpha+\alpha$ elastic scattering. Although the phase shifts given by the present work look good, the description of the $2\alpha$ disintegration threshold needs to be improved. It is shown in Refs.~\cite{Varga:1995dm,Barnea:1999be} that, the exact binding energy of ${}^{4}$He given by the Minnesota force should be around 30 MeV. Therefore, the exact $2\alpha$ disintegration threshold should be around $-60$ MeV, which is much smaller than the value of $-48.5668$ MeV given by the cluster model. One possible way to improve this situations could be combining our theoretical framework with the antisymmetrized molecular dynamics (AMD) \cite{Kanada-Enyo:2001yji,Kanada-Enyo:2012yif} $+$ real-time evolution method (REM) \cite{Imai:2017irh,Imai:2018lww}. The work in this direction is currently under preparation and may be discussed in future publications. 
It is also interesting to extend the analysis here to heavier nuclei such as ${}^{12}$C, ${}^{16}$O, and ${}^{20}$Ne. It is particularly interesting to study the resonant and reaction properties of the Hoyle and high-lying Hoyle-like states \cite{Funaki:2015uya,Schuck:2017jtw,Tohsaki:2017hen,Freer:2017gip,Zhou:2019cjz,Tohsaki:2001an,Bai:2018gqt,Yamada:2003cz,Bai:2018,Barbui:2018sqy,Bishop:2019tqd} with explicit treatments of the asymptotic boundary conditions. Recently, Refs.~\cite{Vadas:2015xor,Schuetrumpf:2017rwv,Wang:2019wbx,Wang:2020} suggest that $\alpha$-cluster structures could be important in understanding some fusion reactions of light nuclei. A combination of our formalism with an imaginary optical potential may also allow microscopic studies of these processes \cite{Descouvemont:1989bdy}. Extending the analysis to the up-right corner of the nuclide chart could be another working direction, where the medium-mass and heavy nuclei such as ${}^{104}$Te and ${}^{212}$Po are known to have rich cluster structures \cite{Auranen:2018usv,Bai:2018giu,Bai:2018hbe,Souza:2019eee,Xiao:2019ywj}. Recently, inspired by the non-localized cluster model, the quartetting wave function approach and the quartet model \cite{Ropke:2014wsa,Xu:2015pvv,Xu:2017vyt,Ropke:2017qck,Bai:2018bjl,Bai:2019jmv,Yang:2019oze} are proposed to study $\alpha$ clustering in heavy nuclei such as ${}^{212}$Po. It is tempted to study the nuclear reactions of medium-mass and heavy nuclei in a similar approach \cite{Cheng:2019kaw,Liu:2019ylg,Liang:2018wzc}.

\begin{acknowledgments} 

D.~B.~would like to thank Mengjiao Lyu for helpful communications on effective nucleon-nucleon interactions, Emiko Hiyama for useful comments during his stay at Department of Physics, Kyushu University, and Bo Zhou for introducing REM to him. 
D.~B.~would also like to thank the anonymous referee for his/her kind guidance.
This work is supported by the National Natural Science Foundation of China (Grant No.~11905103, 11535004, 11947211, 11975167, 11761161001, 11375086, 11565010, and 11881240623), by the National Key R\&D Program of China (Contract No.~2018YFA0404403, 2016YFE0129300), by the Science and Technology Development Fund of Macau under Grant No.~008/2017/AFJ, and by China Postdoctoral Science Foundation (Grant No.~2019M660095).

\end{acknowledgments}

\end{document}